\documentclass[11pt]{article}
\usepackage[T1]{fontenc}
\usepackage[utf8]{inputenc}
\IfFileExists{lmodern.sty}{\usepackage{lmodern}}{}
\usepackage{textcomp}
\usepackage{newunicodechar}
\usepackage[margin=1in]{geometry}
\usepackage{amsmath,amssymb}
\usepackage{graphicx}
\usepackage{array,calc,booktabs,longtable}
\usepackage[numbers,sort&compress]{natbib}
\usepackage[hidelinks]{hyperref}
\newunicodechar{°}{\ensuremath{^{\circ}}}
\newunicodechar{×}{\ensuremath{\times}}
\newunicodechar{Δ}{\ensuremath{\Delta}}
\newunicodechar{Λ}{\ensuremath{\Lambda}}
\newunicodechar{Ω}{\ensuremath{\Omega}}
\newunicodechar{α}{\ensuremath{\alpha}}
\newunicodechar{γ}{\ensuremath{\gamma}}
\newunicodechar{δ}{\ensuremath{\delta}}
\newunicodechar{ε}{\ensuremath{\varepsilon}}
\newunicodechar{λ}{\ensuremath{\lambda}}
\newunicodechar{μ}{\ensuremath{\mu}}
\newunicodechar{ρ}{\ensuremath{\rho}}
\newunicodechar{τ}{\ensuremath{\tau}}
\newunicodechar{φ}{\ensuremath{\varphi}}
\newunicodechar{ω}{\ensuremath{\omega}}
\newunicodechar{′}{\ensuremath{'}}
\newunicodechar{→}{\ensuremath{\rightarrow}}
\newunicodechar{−}{\ensuremath{-}}
\newunicodechar{≈}{\ensuremath{\approx}}
\newunicodechar{⟩}{\ensuremath{\rangle}}

\title{Noise Correlation Spectroscopy: A Review}
\author{Gombojav O. Ariunbold$^{1,2,3}$\thanks{Correspondence: \href{mailto:ag2372@msstate.edu}{ag2372@msstate.edu}}\\[6pt]
\small $^{1}$\,Department of Physics and Astronomy, Mississippi State University, Mississippi State, MS 39762, USA\\
\small $^{2}$\,Physics Department, Wesleyan University, Middletown, CT 06459, USA\\
\small $^{3}$\,Department of Physics and Astronomy, Texas A\&M University, College Station, TX 77843, USA}
\date{}

\begin{document}
\maketitle

\begin{abstract}
The frequency noise of an ordinary diode laser, often a nuisance in
spectroscopy, can be used to measure atomic coherence. When atoms are
prepared in a coherent superposition of two ground states, each random
shift in laser frequency changes how much light they absorb, so the
transmitted light carries a record of that coherence. Correlating the
intensity fluctuations of two fields, each linking one ground state to a
shared excited state, reads this record out, a technique known as noise
correlation spectroscopy. The correlation signal is narrower than the
transparency window the same fields create and switches sign as their
frequency difference is tuned. Its width is set by the rate at which the
atoms lose coherence and, over a wide range, does not grow with laser
power. Over two decades the method has spread from warm vapors to cold
atoms, to multi-field schemes, and to diamond and rare-earth-doped
crystals. This review, to the author's knowledge the first, brings
together its theory, experiments and applications, and examines its
limits, the challenges that remain, and the opportunities that could
make it a standard spectroscopic tool.

\end{abstract}

\section{Introduction}\label{sec:1}

In a coherently prepared atomic medium, laser phase noise can serve as a
measuring tool. A free-running external-cavity diode laser has
negligible amplitude noise but substantial phase noise: its
instantaneous frequency wanders, and in free space a photodiode cannot
detect this wandering. When the beam passes through an atomic vapor
whose absorption changes steeply with frequency, however, each excursion
of the instantaneous frequency becomes an excursion of the absorption.
The beam then leaves the vapor carrying intensity noise that it did not
have on entry. The medium thus transfers its own absorptive and
dispersive structure onto the noise spectrum of the light, so that
measuring the noise measures the medium. Yabuzaki, Mitsui and Tanaka
first demonstrated this explicitly, in the intensity-noise spectrum of a
diode laser transmitted through a resonant vapor \cite{r1}. Walser and
Zoller developed the theory of polarization fluctuations induced by
laser noise \cite{r2}, and Martinelli and co-workers applied the same idea
to a noise-based study of nonlinear magneto-optical resonances \cite{r3}.
Together, these three works established noise spectroscopy as a
technique.

A new technique emerges when two optical modes are used instead of one.
Two beams derived from the same noisy laser and addressing two different
transitions of the same atomic system inherit identical phase noise, but
the medium does not respond to them identically. The physical
information then resides not in the noise power of either beam alone but
in the cross-correlation between their fluctuations. This
cross-correlation reports on the \emph{relative} behavior of two atomic
coherences that belong to one quantum system. It can be positive or
negative, and its sign is fixed by the internal structure of the medium.
The conditions under which the sign changes are a central subject of
this review. The technique built on this cross-correlation is referred
to here as noise correlation spectroscopy. The name joins two terms
already established in the literature. Noise spectroscopy denotes the
earlier single-beam measurements, in which the intensity-noise spectrum
of a laser transmitted through a resonant vapor carries spectroscopic
information \cite{r1,r2,r3}. Correlation spectroscopy is the term adopted for
two-mode correlation measurements of EIT resonances, as in the
power-broadening-free correlation spectroscopy of cold atoms \cite{r4,r5}.
The combined term has also been used for the same observable \cite{r6}. It
states both the source of the signal, the laser's own phase noise, and
the quantity measured, the correlation between two noise records. It
also distinguishes the technique from fluorescence correlation
spectroscopy and from photon-counting (Hanbury Brown--Twiss) correlation
measurements, which probe different physical processes.

The idea is realized most cleanly in the three-level Λ configuration, in
which two optical fields couple an excited state to two long-lived
ground sublevels. Coherent phenomena demonstrated in this system, namely
coherent population trapping (CPT) \cite{r7} and electromagnetically
induced transparency (EIT) \cite{r8,r9,r10}, underlie a large body of work on
slow light, quantum memory \cite{r11}, magnetometry \cite{r12} and
chip-scale atomic clocks \cite{r13}. The defining feature of the Λ system
is a coherence between the two ground sublevels whose lifetime is
limited not by spontaneous emission but by collisions, by the transit of
atoms through the beam and by stray magnetic fields. Its decoherence
rate can therefore be far smaller than the optical linewidth. This
coherence is long-lived, and a magnetic field can continuously tune the
two-photon detuning that controls it (the mismatch between the frequency
difference of the two fields and the splitting of the two ground
sublevels). The Λ system therefore gives the experimenter a single
control parameter. As the field is varied, the correlation between the
two optical modes passes from one sign to the other.

Sautenkov, Rostovtsev and Scully were the first to demonstrate this
switching experimentally \cite{r14}. In a coherently prepared rubidium
vapor, the intensity fluctuations of two circularly polarized
transmitted beams were strongly correlated at zero magnetic field and
became strongly \emph{anti}correlated once a small longitudinal field
was applied. The switch is sharp and is controlled by a single
parameter, and the resonance in the correlation signal is about four
times narrower than the EIT transmission resonance measured on the same
sample at the same optical power. Intensity correlations between EIT
fields had been seen before. Garrido Alzar and co-workers measured
positive correlations between pump and probe fields from two independent
lasers on EIT resonance \cite{r15}. The noise-spectrum measurements of
Martinelli and co-workers showed anticorrelated noise between the two
circular polarization components of a single beam near zero magnetic
field \cite{r3}. The switching of the correlation itself, from strong
correlation to strong anticorrelation under the control of the field,
was first measured in Ref.~\cite{r14}, and this observation is the
starting point of the present review.

The first theoretical explanation of this switching was given by
Ariunbold, Sautenkov, Rostovtsev and Scully in a preprint posted in
March 2006, three months after the experiment was published, and later
published in full \cite{r16}. The stochastic model and the numerical
calculations in that work were developed by its first author, who is
also the author of this review. This treatment gives the driving field a
finite bandwidth, so that the laser phase enters the equations of motion
as a multiplicative random process. It identifies the transmitted
intensity fluctuations with the imaginary parts of the two optical
coherences, which means that the measured correlation function is, at
its core, a correlation function of atomic quantities. It traces the
sign change to an antisymmetric fluctuation of the two ground-state
populations that is present only away from two-photon resonance.
Finally, it predicts a switching curve with a universal shape whose only
scale is the ground-state decoherence rate. Above a modest threshold
this scaled curve does not broaden as the drive power is raised (it
narrows slightly and then levels off). The theory therefore predicted,
before any systematic measurement, that the correlation resonance is
free of power broadening. For the same reason, the comparison with the
measured data needs no fitting parameters, only a single scale factor.
Sections 2--4 and 7 present these four elements (an observable, a
mechanism, a solvable model and a measurement protocol) in the detail
required by the later sections. Because Sections 4 and 7 rely
substantially on the author's own work, readers should weigh the
emphasis accordingly.

Two decades of work have followed the first experimental demonstration
by Sautenkov et al. \cite{r14} and its theoretical explanation by
Ariunbold et al. \cite{r16}. The technique has been refined into a
spectroscopy whose correlation linewidth, as the theory of Ref.~\cite{r16}
predicted, remains insensitive to power broadening over the range tested
\cite{r4}. Its correlation-to-anticorrelation transition has a distinct
spectroscopic linewidth that can fall below the
coherence-lifetime-limited EIT width \cite{r17}, and it resolves spectral
structure that transmission spectroscopy of the same sample cannot
\cite{r18}. It has been extended to cold atoms \cite{r5,r19}, to
electromagnetically induced \emph{absorption} \cite{r20}, to buffer-gas
cells in which collisions set the ground-state decoherence rate
\cite{r21,r22}, and to optical cavities \cite{r23}. The two-mode correlation
has been generalized to three, four and six mixed fields \cite{r24,r25,r26},
to correlations between two \emph{generated} four-wave-mixing (FWM)
signals \cite{r27}, and to a velocity-resolved treatment of these
correlations in hot vapor \cite{r28}. It has also moved beyond alkali
vapors altogether, to nitrogen--vacancy centers in diamond \cite{r29} and
rare-earth-doped crystals \cite{r30}, and most recently to structured
light, in the generation of correlated optical vortices \cite{r31}. Along
the way, the theory has acquired frequency-domain \cite{r32} and Lindblad
master-equation \cite{r33} formulations, and has been extended to include
the laser's residual intensity noise as a second noise channel \cite{r21}.

To the author's knowledge, no review devoted to this technique has
appeared so far. The present one is organized as follows. Section 2
introduces the physical setting: the Λ system and its coherences, the
conversion of phase noise into intensity noise, and the two-mode
correlation function in the time and frequency domains. Section 3
describes the correlation-switching phenomenon and explains its
mechanism. Section 4 presents the stochastic density-matrix treatment
with its analytic and numerical routes, what it predicts for the scaling
of the switching curve, and how far these predictions hold in a real
cell. It also covers the frequency-domain, perturbative and
master-equation approaches developed alongside it. Section 5 describes
the standard apparatus and collects the quantitative benchmarks against
which later work can be compared. Section 6 discusses spectroscopy with
resonances narrower than the natural linewidth, and Section 7 the
measurement of decoherence rates. Section 8 surveys the extensions of
the technique and the new effects they have revealed. Section 9 turns to
applications, Section 10 to how the framework has been taken up in the
wider literature, and Section 11 summarizes the conclusions and the
problems that remain open.

This review concerns classical intensity-noise correlations produced by
a noisy classical drive. Quantum correlations of the light itself, such
as squeezing and entanglement of generated fields, arise from a
different mechanism and form a separate subject that is not treated
here.

\section{Physical setting}\label{sec:2}

\subsection{The Λ system, ground-state coherence and coherent preparation}\label{sec:2.1}

Throughout this review the system is a three-level Λ configuration
(Figure 1a). It has an excited state \(|a\rangle\) and two ground
sublevels \(|b\rangle\) and \(|c\rangle\), with the transition
\(\left| a\rangle \leftrightarrow \right|b\rangle\) driven by a field of
Rabi frequency \(\Omega_{1}\) and the transition
\(\left| a\rangle \leftrightarrow \right|c\rangle\) by a field of Rabi
frequency \(\Omega_{2}\). (The Rabi frequency measures the strength of
the coupling between the atom and the field.) In the experiments
reviewed here, \(|b\rangle\) and \(|c\rangle\) are Zeeman sublevels of a
single hyperfine ground state, typically \(5S_{1/2}(F = 2)\) of
\textsuperscript{87}Rb, and the two fields are the two counter-rotating
circular components of a single beam. Thus \(\Omega_{1}\) and
\(\Omega_{2}\) originate from one laser and carry identical phase noise.

Three rates set the timescales of the system. The first two are
radiative. The optical coherences \(\rho_{ab}\) and \(\rho_{ac}\) decay
at the transverse optical rate \(\gamma \sim \Gamma_{r}/2\), of the
order of several MHz for the rubidium D\textsubscript{1} line, and the
excited-state population decays at \(\Gamma_{r}\). The third rate is of
a different kind. The ground-state coherence \(\rho_{bc}\) decays at a
rate \(\gamma_{bc}\) (written \(\gamma_{3}\) in much of the theoretical
literature) that is set by transit time, residual field gradients and
collisions, not by any radiative process. Thus \(\gamma_{bc}\) depends
entirely on the cell. In a buffer-gas or wall-coated cell it can lie
orders of magnitude below the optical linewidth. In an unbuffered
room-temperature cell with a millimeter-scale beam, transit time alone
sets a rate of a few times 10\textsuperscript{5} s\textsuperscript{−1}
(a 0.1 cm beam crossed at a few hundred m/s gives v/d ≈ 3 ×
10\textsuperscript{5} s\textsuperscript{−1}, roughly 50 kHz in ordinary
frequency). The value extracted for the vapor of Section 5 is of order 1
MHz, the same order as the radiative decay rate \(\gamma\). Because
\(\gamma_{bc}\) is set by the sample and not by the drive, it is the key
quantity to be measured.

\begin{figure}[!htbp]
\centering
\includegraphics[width=0.92\linewidth]{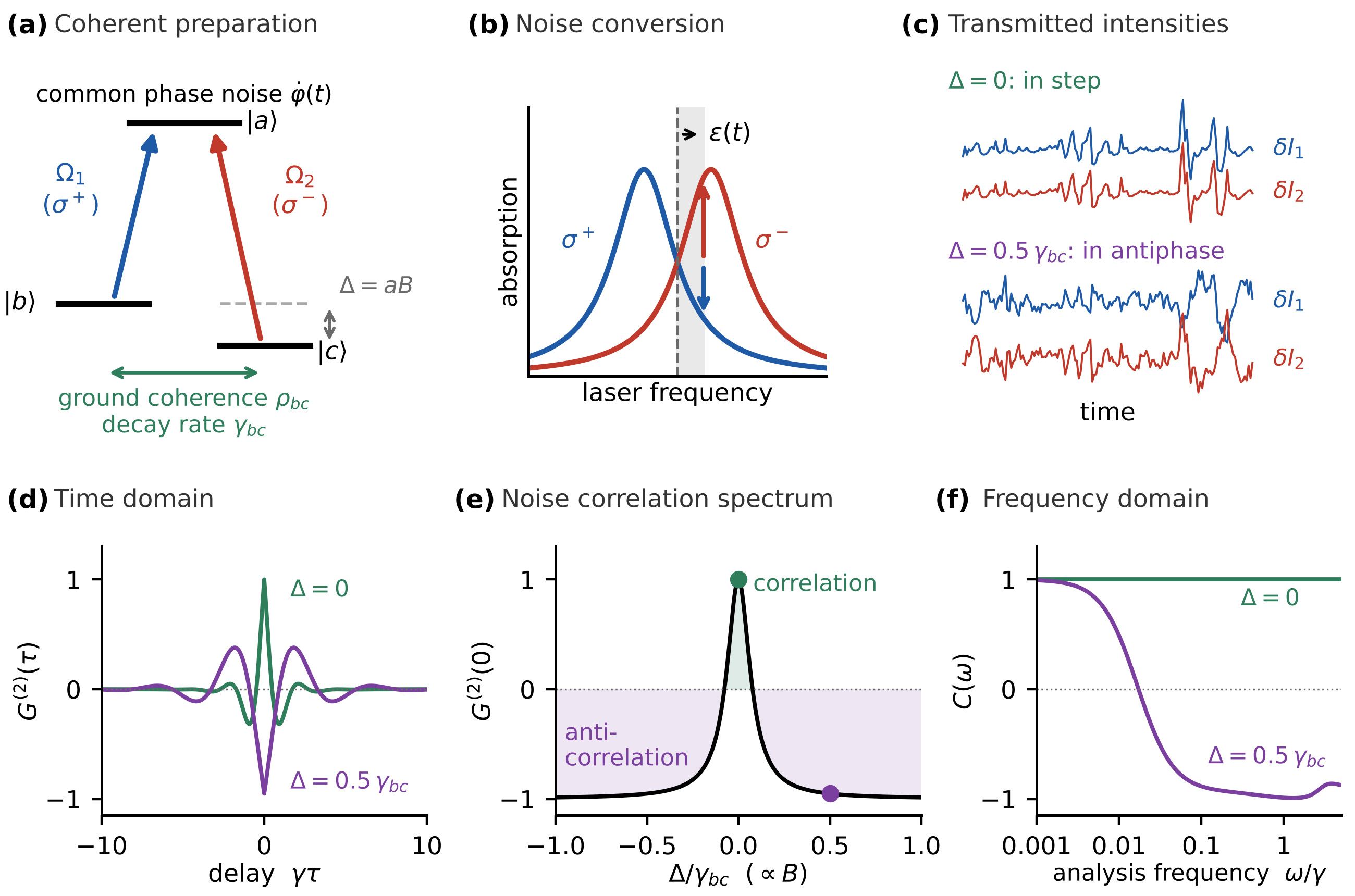}
\caption{Noise correlation spectroscopy at a glance. (\textbf{a}) Coherent preparation: a Λ system with excited state \textbar a⟩ and two Zeeman ground sublevels \textbar b⟩ and \textbar c⟩, driven by the two circular components of one laser, so that both transitions carry the same phase noise \ensuremath{\dot\varphi}(t); a longitudinal field B sets the two-photon detuning Δ = aB, and the ground-state coherence ρ\textsubscript{bc} decays at the rate γ\textsubscript{bc}. (\textbf{b}) Noise conversion: when the sublevels are split, a common excursion ε(t) of the laser frequency moves one transition toward resonance and the other away from it, so one beam is absorbed more (red arrow) and the other less (blue arrow). (\textbf{c}) Transmitted intensity fluctuations δI\textsubscript{1} and δI\textsubscript{2}: in step at Δ = 0 and in antiphase at Δ = 0.5γ\textsubscript{bc}. (\textbf{d}) Time domain: their normalized cross-correlation G\textsuperscript{(2)}(τ), a correlation peak (+1.00) at Δ = 0 and an anticorrelation dip (−0.95) at Δ = 0.5γ\textsubscript{bc}. (\textbf{e}) The noise correlation spectrum, G\textsuperscript{(2)}(0) as the field B (and hence Δ) is scanned: a narrow correlation resonance whose width is set by γ\textsubscript{bc}, not by the laser power, flanked by anticorrelation; the dots mark the two cases of (c) and (d). (\textbf{f}) Frequency domain: the frequency-resolved correlation C(ω) = S\textsubscript{12}/(S\textsubscript{11}S\textsubscript{22})\textsuperscript{1/2} of Section 2.4 against analysis frequency; it stays at +1 for Δ = 0, while for Δ = 0.5γ\textsubscript{bc} the two noises are correlated only at the lowest frequencies and anticorrelated above about 0.17γ\textsubscript{bc}. Panels (a) and (b) are schematic; panels (c)--(f) are computed from the stochastic model of Section 4.1 (Ω\textsubscript{1} = Ω\textsubscript{2} = γ, γ\textsubscript{bc} = 0.1γ, D = 0.01γ), the same model used for Figure 2; (c) shows one Monte Carlo trajectory per case.}\label{fig:1}
\end{figure}

Driving both transitions coherently prepares the atoms in a
superposition of \(|b\rangle\) and \(|c\rangle\) that is decoupled from
the light, the dark state of coherent population trapping \cite{r7}, so
that the medium becomes transparent on two-photon resonance \cite{r9,r10}.
The width of this transparency window is set jointly by \(\gamma_{bc}\)
and by the drive strength. The drive term is the main drawback of
conventional EIT spectroscopy: increasing the drive to improve the
signal broadens the resonance as \(\Omega^{2}/\gamma\). Section 6
explains how the correlation observable avoids this limitation.

A longitudinal magnetic field \(B\) lifts the Zeeman degeneracy and
detunes the two-photon resonance by

\[\Delta = aB,\quad\quad a = \frac{\mu_{B}}{h}\, g\,\left( m_{2} - m_{1} \right),\]

which, for the \(5S_{1/2}(F = 2)\) ground state of
\textsuperscript{87}Rb with \(g = 0.5\) and \(\mu_{B}/h = 1.4\) MHz/G,
gives \(a = 1.4\) MHz/G for the two-photon detuning between the
\(m = \pm 1\) sublevels addressed by the two circular components
\cite{r16}. The Zeeman shift of a single sublevel is half this value, 0.7
MHz/G. Both numbers appear side by side in the literature and are
consistent, but the factor of two matters when a decoherence rate is
extracted from a magnetic-field scale. The two must therefore be kept
distinct. Because \(B\) maps linearly onto \(\Delta\) with a known
coefficient, the magnetic field serves as a convenient control
parameter. A two-layer magnetic shield suppresses the ambient field
enough that the applied field is, to a good approximation, the only
source of two-photon detuning. How good this approximation actually is
turns out to be the main limitation of the quantitative protocol of
Section 7, as discussed in Section 4.3.

\subsection{Conversion of phase noise into intensity noise}\label{sec:2.2}

The laser field is modelled as
\(E(t) = E_{0}\exp\left\lbrack - i\omega t - i\varphi(t) \right\rbrack + c.c.\)
with a fluctuating phase. In the standard phase-diffusion (Wiener--Lévy)
model, the instantaneous frequency deviation \(\dot{\varphi}(t)\) is
Gaussian white noise \cite{r16},

\[\langle\dot{\varphi}(t)\rangle = 0,\quad\quad\langle\dot{\varphi}(t)\dot{\varphi}(t')\rangle = 2D\,\delta(t - t'),\]

corresponding to a Lorentzian laser spectrum with a full width of
\(D/\pi\) Hz. The same phase-diffusion process underlies the
Schawlow--Townes-type linewidth of a semiconductor laser. For a typical
free-running external-cavity diode laser this linewidth is of order 1
MHz: comparable to the optical linewidth and much larger than
\(\gamma_{bc}\).

Note that \(\dot{\varphi}(t)\) enters the equations of motion
\emph{multiplicatively}. In a frame rotating with the instantaneous
laser phase, the coherence equations acquire a term
\(i\dot{\varphi}(t)\rho\). Unlike additive noise, multiplicative noise
cannot be averaged separately from the dynamics, and this is what makes
the theoretical treatment of Section 4 difficult. Physically, the medium
converts phase noise into amplitude noise. The conversion is strongest
where the absorption changes most steeply with frequency.

This conversion has two consequences.

First, the noise of the transmitted light is \emph{generated by the
medium}. It is not the laser's own amplitude noise passed through
unchanged but a signal created by the interaction. Its spectrum and
correlations therefore carry information about the medium. This holds
provided that the laser's own residual intensity noise (RIN) is small
enough to neglect; when it is not, the departure is informative, as
Section 8.3 shows.

Second, because both modes come from the same laser, they experience the
\emph{same} frequency excursion at every instant. Whether the resulting
intensity fluctuations are correlated or anticorrelated therefore
depends only on whether the two transitions respond to the shared
excursion in the same way or in opposite ways. The phenomenon thus
reduces to the relative sign of two susceptibilities, which is the
subject of Section 3.

\subsection{The two-mode intensity--intensity noise correlation function}\label{sec:2.3}

The central observable is the normalized second-order cross-correlation
of the intensity \emph{fluctuations} of two distinct optical modes
\cite{r14,r16},

\[G^{(2)}(\tau) = \frac{\langle\delta I_{1}(t)\,\delta I_{2}(t + \tau)\rangle}{\left\lbrack \langle\delta I_{1}(t)^{2}\rangle\langle\delta I_{2}(t + \tau)^{2}\rangle \right\rbrack^{1/2}},\]

where \(\delta Q(t) = Q(t) - \bar{Q}(t)\) is the deviation from the
running mean, and the average is taken over a record of finite duration
\(T\) (10 μs in the original measurements) \cite{r14,r16}.

The quantity has three important properties. First, it is a
cross-correlation between the two counter-rotating circular
polarizations, not the autocorrelation of a single beam. This
distinguishes it from single-beam noise spectroscopy \cite{r1,r2}: because
the two modes address different Zeeman coherences of the same Λ system,
their cross-correlation reports on the relative behavior of two
coherences. Common-mode contributions, such as laser RIN, mechanical
drift and fluctuations in detector gain, enter both channels with the
same sign and therefore push \(G^{(2)}\) toward positive values.
Common-mode noise therefore cannot produce an anticorrelation, and a
measured negative value must originate in the medium. On the positive
side the two contributions are not so cleanly separated, and Section 8.3
describes a measurement in which they were not.

Second, the quantity is normalized, so \(G^{(2)}\) is bounded and
dimensionless, and its values can in principle be compared across
experiments, media and optical powers. This would allow the benchmarks
of Section 5.2 to be transferred between laboratories.

Third, the quantity is sign-resolved: \(G^{(2)}(0) > 0\) indicates
correlation and \(G^{(2)}(0) < 0\) anticorrelation, and the sign has
physical meaning. The anticorrelated branch is the signature of the
effect, and the change of sign is the phenomenon examined in Section 3.

On notation: the symbol \(g^{(2)}\) conventionally denotes the
normalized second-order coherence function of quantum optics,
\(g^{(2)}(\tau) = \langle:\widehat{I}(t)\widehat{I}(t + \tau):\rangle/\langle\widehat{I}\rangle^{2}\),
whose value at zero delay distinguishes bunched, coherent and
antibunched light. That is not the quantity defined above. The quantity
here is a \emph{classical} correlation coefficient of intensity
fluctuations, bounded between \(- 1\) and \(+ 1\); a value of \(- 0.9\)
indicates strong classical anticorrelation and says nothing about
sub-Poissonian photon statistics. Several papers nevertheless write
\(g_{2}(0)\) for the classical quantity, and the reader must infer which
is meant from the definition given, not from the symbol.

\subsection{Time domain and frequency domain}\label{sec:2.4}

Two related descriptions are used in the literature. One works in the
time domain with \(G^{(2)}(\tau)\) computed from digitized photocurrent
records. The other works in the frequency domain with the symmetrized
spectral densities of the intensity fluctuations \cite{r5},

\[S_{ij}(\omega) = \frac{1}{4\pi}\int_{- \infty}^{\infty}d\tau\, e^{- i\omega\tau}\left\lbrack \langle\delta I_{i}(t)\,\delta I_{j}(t + \tau)\rangle + \langle\delta I_{j}(t)\,\delta I_{i}(t + \tau)\rangle \right\rbrack,\]

and the frequency-resolved correlation is

\[C(\omega) = \frac{S_{12}(\omega)}{\sqrt{S_{11}(\omega)\, S_{22}(\omega)}},\]

which is measured directly with a spectrum analyzer \cite{r5,r32}. The two
descriptions are closely related but not identical: it is the
cross-spectral density \emph{S}\textsubscript{12}(ω), not the normalized
\emph{C}(ω), that is the Fourier transform of the symmetrized two-time
intensity correlation, and each observable is normalized in its own
domain. The choice between them is therefore largely practical.

The zero-delay value \(G^{(2)}(0)\) integrates over the entire detection
bandwidth and yields a single number for each setting of the control
parameter; its dependence on magnetic field is the switching curve, also
called here the noise correlation spectrum. The frequency-resolved
correlation \(C(\omega)\) instead resolves the correlation by noise
frequency, separating timescales that \(G^{(2)}(0)\) combines:
ground-state relaxation appears at low frequencies, optical pumping and
Rabi dynamics at higher ones. Sideband resonances that are invisible in
\(G^{(2)}(0)\) appear as structure in \(C(\omega)\) \cite{r5}, and the
frequency-domain formulation lends itself to a perturbative analytic
treatment that the time-domain approach has so far not allowed \cite{r32}.

Flórez and co-workers make this distinction explicit \cite{r5}: the
treatments of Refs.~\cite{r16,r19} work in the time domain with
\emph{G}\textsuperscript{(2)}(τ), whereas Refs.~\cite{r5,r32} work in the
frequency domain with \emph{C}(ω). Which observable is preferable in
which regime, and how the parameters of one map onto those of the other,
remain open questions.

\section{Correlation switching and its mechanism}\label{sec:3}

\subsection{The observation}\label{sec:3.1}

Sautenkov, Rostovtsev and Scully reported the phenomenon in a coherently
prepared rubidium vapor \cite{r14}. Two circularly polarized beams from a
single external-cavity diode laser, tuned to the \textsuperscript{87}Rb
D\textsubscript{1} line, pass through a magnetically shielded 7.5 cm
cell and are then separated onto two fast photodiodes. At zero magnetic
field the two beams fluctuate together, with a strong positive
correlation, \(G^{(2)}(0) \approx + 0.9\). When a longitudinal field of
a few tenths of a Gauss is applied, the correlation passes through zero
to a strong anticorrelation, \(G^{(2)}(0) \approx - 0.9\) at
\(B = - 0.47\) G. The resonance of the correlation as a function of
field has a full width of 0.24 G, compared with an EIT transmission
width of 0.85 G measured on the same sample at the same total power of 1
mW. The correlation resonance is about four times narrower. A second
data set in the same paper, taken at 0.25 mW per beam, narrows the
correlation resonance to 0.16 G, with a correspondingly narrower EIT
window. The authors report the same roughly fourfold ratio at this lower
power but do not extend the comparison beyond these two points.

Two control measurements indicate that the effect originates in the
medium rather than in the optics. When the two beams are displaced
transversely by 0.3 cm (three beam diameters), the correlation persists,
weaker but clearly present (\(+ 0.7\) against a background of
\(\approx 0.3\)), which rules out common-mode optical overlap as its
source. In the second control, the two detection paths are matched in
length, so that a correlation measured at \(\tau = 0\) reflects the
medium and not a relative optical delay.

The 2005 paper does not develop a theory of its own. It states only that
the results were interpreted with a theoretical approach similar to
earlier work on correlated-emission lasers (Section 3.3), and the word
\emph{decoherence} does not appear in it. The group went on to study
power spectra and correlations of intensity fluctuations under EIT and
in nonlinear magneto-optical rotation \cite{r34,r35}, and a companion
measurement of intensity correlations in coherently prepared Rb vapor in
a magnetic field followed \cite{r36}. The mechanism behind the sign
change, and the link between the switching field and the ground-state
decoherence rate, come from the stochastic treatment described next.

\subsection{Intensity fluctuations as a readout of atomic coherences}\label{sec:3.2}

The first theoretical step connects a measurement on the light to a
property of the atoms \cite{r16}. For an optically thin medium with weak
absorption, the transmitted Rabi frequencies are, to first order in the
propagation length \(L\),

\[\Omega_{1}^{out} = \Omega_{1} + i\,\eta_{1}L\,\rho_{ab}(t),\quad\quad\Omega_{2}^{out} = \Omega_{2} + i\,\eta_{2}L\,\rho_{ac}(t),\]

with coupling constants
\(\eta_{1,2} = 3\lambda^{2}N\Gamma_{r1,2}/8\pi\), valid for
\(\eta L \ll \Gamma_{r}\). Taking intensities and keeping first-order
terms gives

\[\delta I_{1} = \Omega_{1}\eta_{1}L\text{\,Im}\{\delta\rho_{ab}(t)\},\quad\quad\delta I_{2} = \Omega_{2}\eta_{2}L\text{\,Im}\{\delta\rho_{ac}(t)\}.\]

Here Im denotes the imaginary part. Up to positive constants that cancel
in the normalization, the measured \(G^{(2)}(\tau)\) \emph{is} the
correlation function of the imaginary parts of two optical coherences. A
noise measurement on the transmitted light is therefore a direct readout
of fluctuations in the atomic coherences. The detection need not be
quantum-limited or interferometric. Two fast photodiodes and a digitizer
suffice, because the medium performs the conversion. The detection
bandwidth and record length must, however, reach down to the
ground-state timescale; Section 5.1 gives the values used.

Two further points matter later. First, the imaginary part is the
absorptive response, so the observable follows the dissipative rather
than the dispersive part of the susceptibility. This becomes relevant
when the technique is applied to processes such as four-wave mixing,
where the amplitude of a generated field is measured instead of the
attenuation of a transmitted one. Second, the original derivation
neglects Doppler averaging, on the grounds that the two-photon detuning
of a Λ system driven by two co-propagating fields is Doppler-free to
first order. This is an explicit approximation; the four-wave-mixing
work of Section 8.8 treats it with an explicit velocity average.

\subsection{Why the beams are correlated on two-photon resonance}\label{sec:3.3}

Let us consider first \(\Delta = 0\). The configuration is then
symmetric: the two ground sublevels are equally populated,
\(\rho_{bb} = \rho_{cc}\), and the ground-state coherence \(\rho_{bc}\)
is real. The equations of motion for the two optical coherences become
identical, \(\rho_{ba}(t) = \rho_{ca}(t)\), and the two modes carry
identical noise.

Both transitions experience the same instantaneous detuning excursion
and, because the configuration is symmetric, respond to it in the same
way: both beams brighten and darken together, and
\(G^{(2)}(0) \rightarrow + 1\) (measured: \(+ 0.9\)). The original
treatment draws an analogy with the correlated-emission laser
\cite{r37,r38}, in which the relative phase of two fields sharing an atomic
system fluctuates less than the phase of either field alone. Here, the
two intensities share a fluctuation because they share the atomic system
that generates it. The comparison is only an analogy.
Correlated-emission lasing concerns relative phase, whereas the present
effect concerns intensity.

\subsection{Why the sign flips: antisymmetric ground-state population fluctuations}\label{sec:3.4}

Let us now consider \(\Delta \neq 0\). Because the total ground-state
population is conserved, \(\delta\rho_{bb} + \delta\rho_{cc} = 0\), the
populations can be written as

\[\rho_{bb}(t) = c + f(t),\quad\quad\rho_{cc}(t) = c - f(t),\]

with \(f(t)\) real and non-zero only because \(\Delta \neq 0\). This
antisymmetric population fluctuation enters the two polarization
equations with opposite signs,

\[\rho_{ba} \sim + F(t) + c_{1},\quad\quad\rho_{ca} \sim - F(t) + c_{2},\]

and the cross-correlation changes sign with it.

Figure 1b,c illustrates this picture. Once the Zeeman degeneracy is
lifted, a common frequency excursion moves the two transitions in
\emph{opposite} directions relative to their resonances: bringing the
\(\sigma^{+}\) transition closer to resonance moves the \(\sigma^{-}\)
transition further away. The two beams therefore brighten and darken in
antiphase, and the anticorrelation is carried by \(f(t)\), the
antisymmetric fluctuation of the ground-state populations. This is a
\emph{ground-state} effect: its sign is controlled by how population is
redistributed between two long-lived sublevels. That is why the relevant
field scale is \(\gamma_{bc}\) rather than the optical linewidth, and
why the correlation resonance can be narrower than the transparency
window of the same sample. Figure 1d--f shows the result: a correlation
peak in G\textsuperscript{(2)}(τ) at Δ = 0, an anticorrelation dip once
the sublevels are split, a noise correlation spectrum, as the field is
scanned, whose central resonance has a width set by γ\textsubscript{bc},
and the same behavior in the frequency domain.

Nothing in this mechanism requires the medium to be optically thick, the
drive to be weak, or the two fields to be quantum-mechanically
correlated. The effect is a classical stochastic response of a
coherently prepared medium, and it has since been observed in hot vapor,
cold atoms, buffer-gas cells and solid-state hosts. This breadth is
consistent with the generality of the argument, with one caution. As
Section 8.9 argues, the solid-state cases share only the definition of
the observable and not this particular mechanism. They should not be
taken as evidence that this specific analysis is the most general one.

\section{Theoretical frameworks}\label{sec:4}

Several theoretical approaches have been applied to the phenomenon. They
differ in their assumptions and in the quantities they can calculate.

\subsection{Stochastic density-matrix treatment of a finite-bandwidth drive}\label{sec:4.1}

The framework adopted in most of the later literature treats the driving
field as having a finite bandwidth, with the laser phase entering the
density-matrix equations of motion as a multiplicative random process.
It was introduced for this problem by the present author and his
collaborators \cite{r16}. Writing the density-matrix elements as a vector
\(C\), the equations take the form

\[\frac{dC}{dt} = M_{0}C + i\dot{\varphi}(t)\, MC,\]

where \(M_{0}\) is the deterministic Liouvillian (the operator that
generates the noise-free evolution, including relaxation) and \(M\) is
the matrix through which the phase noise enters.

Two routes lead on from here. The \emph{analytic route} performs a
functional stochastic average over the multiplicative white noise,
following Wódkiewicz's stochastic-decoupling method \cite{r39}, and yields
a closed equation for the averaged vector,

\[\frac{d\langle C\rangle}{dt} = M_{0}\langle C\rangle - \frac{G}{2}M^{2}\langle C\rangle,\quad\quad G = 2D.\]

This is the finite-bandwidth generalization of the earlier treatment
with a monochromatic drive. The noise contributes a term quadratic in
the coupling matrix. A broader laser produces stronger effective damping
of precisely those coherences through which the noise couples. Because
this damping is not uniform across the density matrix of the Λ system,
the non-uniformity produces the asymmetry described in Section 3.4.

The \emph{numerical route} solves the stochastic equations directly by
Monte Carlo simulation, that is, by averaging over many randomly
generated realizations of the noise. Ornstein--Uhlenbeck (colored)
noise,
\(\langle\xi(t)\xi(t')\rangle = Y\Lambda_{L}\exp\left( - \Lambda_{L}|t - t'| \right)\)
\cite{r40}, is brought close to the white-noise limit by making
\(1/\Lambda_{L}\) small. The noise realizations are generated with the
Box--Muller algorithm \cite{r41} and the fast integral noise generator of
Fox, Gatland, Roy and Vemuri \cite{r42}, integrated with the
Euler--Maruyama scheme, and averaged over about \(10^{3}\) trajectories.
The numerical route is the primary tool here, because the observable is
a two-time correlation function of fluctuations rather than a mean
value: the analytic average gives only the mean dynamics, whereas the
correlation function requires the individual trajectories.

Using colored noise taken to the white-noise limit, instead of white
noise from the outset, has a further advantage. Keeping \(\Lambda_{L}\)
finite makes the laser's coherence time an explicit, adjustable
parameter. The same method should in principle extend to lasers whose
linewidth is not large compared with the atomic response. Whether it
does remains untested: the author has found no calculation in this line
carried out away from the near-white-noise limit, and every experiment
reviewed here uses a free-running diode laser.

\subsection{What the model predicts: scaling of the switching curve}\label{sec:4.2}

Solved numerically over a range of parameters, the treatment of Section
4.1 determines the shape of the switching curve and how its width
depends on the parameters of the medium and of the drive \cite{r16}. Three
results follow.

\textbf{The width follows the ground-state decoherence rate.} When
\(G^{(2)}(0)\) is computed as a function of both the two-photon detuning
\(\Delta\) and the ground-state dephasing rate \(\gamma_{bc}\), the
profile keeps its shape while its width grows steadily as
\(\gamma_{bc}\) increases over the range
\(\gamma_{bc} \approx 0.01\)--\(0.11\,\gamma\). Over this decade, the
width of the correlation feature is a direct measure of \(\gamma_{bc}\),
and not of the optical linewidth, the population decay rates or (as
Section 6 shows) the drive. No calculations were made outside this
decade.

\textbf{The curves collapse onto a single shape.} When the detuning is
rescaled by the same rate, the computed curves fall on top of one
another: plotted against \(\Delta/\gamma_{bc}\) instead of \(\Delta\),
the curves for different decoherence rates coincide (their zero
crossings agree to within 5\%). Within this range, the switching curve
has a single shape whose only scale is \(\gamma_{bc}\), so a measured
curve carries exactly one piece of information about the medium. The
literature calls this behavior ``universal''; what has actually been
shown is a collapse over one decade of \(\gamma_{bc}\), within one
model.

\textbf{Above a threshold, the curve does not broaden with drive
strength.} At fixed \(\gamma_{bc}\), the curves computed for
\(\Omega_{1} = 0.6,0.8,\ldots,4\,\gamma\) nearly coincide; computed
exactly, the stronger drives give a slightly narrower resonance, and the
curves level off toward the top of the range. Only the two weakest
drives, \(\Omega_{1} = 0.2\) and \(0.4\,\gamma\), stand apart clearly:
the \(0.2\,\gamma\) curve is several times broader in the wings than the
collapsed set. Across the computed range, nearly a factor of seven in
Rabi frequency, the correlation resonance does not broaden with the
drive (its full width falls slightly, from 0.19 to 0.13 of the
decoherence rate). The EIT transmission resonance of the same sample
behaves differently: in the weak-drive limit it broadens as
\(\Omega^{2}/\gamma\), and at the stronger drives computed here grows
roughly in proportion to the Rabi frequency (Figure 2d). Drives above
\(4\,\gamma\) were not computed, because the idealized conditions of the
model no longer hold there. In other words, the model predicts a
correlation resonance that is free of power broadening. Figure 2(a,b)
shows both results, recomputed for this review from the model of Section
4.1. Section 6 discusses what this means for spectroscopy in practice,
and Section 6.2 describes the cold-atom measurement that later tested a
closely related property, the power independence of the
correlation-feature width \cite{r4}.

\textbf{Comparison with the measured curve.} Together, these three
results allow a quantitative comparison with experiment using a single
scale factor. Writing the measured field axis as
\(B = \alpha\,\left( \Delta/\gamma_{bc} \right)\) and matching the
computed universal curve to the correlation-versus-field data of the
room-temperature rubidium experiment \cite{r14} gives \(\alpha \approx 1\)
G and hence, through the Zeeman constant \(a = 1.4\) MHz/G, a
ground-state dephasing rate \(\gamma_{bc}\) of order 1 MHz. This scale
factor is the only adjusted quantity; there are no fitting parameters,
and neither the population decay rates nor the drive strength enter the
comparison. Because the scaled curve does not broaden with drive
strength on the plateau and keeps its shape for every decoherence rate
computed, there is nothing else to adjust.

The agreement is close through the central part of the resonance: the
positive peak, the zero crossing and the approach to anticorrelation on
one side. It deteriorates in the wings, where the measured points lie
systematically above the computed curve and are visibly asymmetric about
\(B = 0\), which the symmetric model cannot reproduce. Section 4.3 takes
this mismatch as its starting point.

\begin{figure}[!htbp]
\centering
\includegraphics[width=0.92\linewidth]{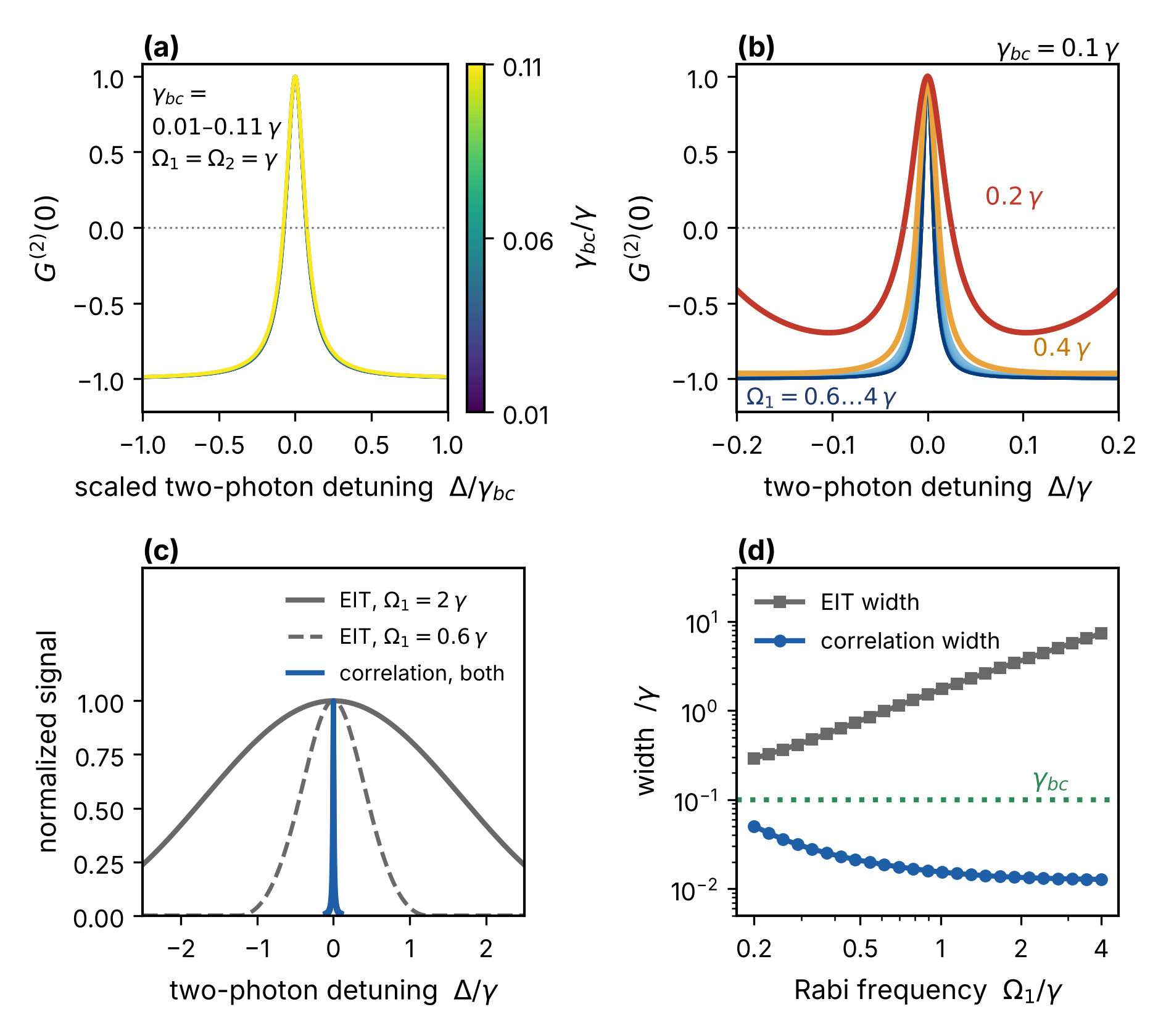}
\caption{The switching curve computed from the stochastic model of Ref.~\cite{r16} (Section 4.1). G\textsuperscript{(2)}(0) is evaluated exactly for white laser phase noise (diffusion coefficient D = 0.01γ, where γ is the optical-coherence decay rate) from the steady-state moments of the stochastic density-matrix equations; a direct Monte Carlo simulation at a representative point agrees to within 0.01. Each curve is evaluated on 8001 detuning points, and the widths in (d) are located by root-finding. (\textbf{a}) Scaling: G\textsuperscript{(2)}(0) against the scaled two-photon detuning Δ/γ\textsubscript{bc} for eleven ground-state decoherence rates, γ\textsubscript{bc} = 0.01, 0.02, \ldots, 0.11γ, at Ω\textsubscript{1} = Ω\textsubscript{2} = γ; once the detuning is scaled by γ\textsubscript{bc}, the curves fall on top of one another. (\textbf{b}) No broadening with drive: G\textsuperscript{(2)}(0) against Δ at γ\textsubscript{bc} = 0.1γ for Rabi frequencies Ω\textsubscript{1} = Ω\textsubscript{2} from 0.2γ to 4γ; the two weakest drives stand apart, while the curves for Ω\textsubscript{1} = 0.6γ to 4γ gather into one narrow bundle. (\textbf{c}) Why the correlation resonance is the sharper observable: the normalized EIT transparency window (gray; dashed Ω\textsubscript{1} = 0.6γ, solid Ω\textsubscript{1} = 2γ) and the correlation peak {[}G\textsuperscript{(2)}(0) + 1{]}/2 (blue) for the same two drives, at γ\textsubscript{bc} = 0.1γ. (\textbf{d}) Full widths at half maximum against drive strength: the EIT width grows steadily with the drive, whereas the width of the correlation peak levels off and stays below γ\textsubscript{bc} (dotted line). Panels (a) and (b) reproduce Figs. 8 and 9 of the 2006 preprint of Ref.~\cite{r16}. The model is Doppler-free and optically thin; the contrast between the two widths is larger here than in a warm vapor.}\label{fig:2}
\end{figure}

\subsection{Limits of the model}\label{sec:4.3}

The three results above are properties of a theoretical model \cite{r16};
a real measurement is more fragile. This insensitivity to the drive
holds only over a limited range of conditions. Four limits are
considered below.

\emph{The drive threshold.} Insensitivity to \(\Omega_{1}\) is not a
general property of the system. It sets in above
\(\Omega_{1} \sim \gamma\) and disappears below it. The failure below
threshold is abrupt, with the \(\Omega_{1} = 0.2\,\gamma\) curve several
times broader in the wings than the collapsed set. An experiment run at
low power to avoid perturbing the sample therefore sits in the regime
where the width does depend on power, the opposite of what the technique
requires. Where this threshold lies in a given cell, in laboratory units
rather than in units of \(\gamma\), has not been measured.

\emph{The upper end.} Nothing in the model excludes a failure of this
insensitivity at high drive as well. Optical pumping out of the
addressed sublevels, light shifts of the Zeeman structure and saturation
of the transition should, on general grounds, reintroduce a power
dependence at some point, and no experiment has yet located it.

\emph{Optical thickness, Doppler broadening and amplitude noise.} The
model assumes an optically thin, Doppler-free medium and a laser without
amplitude noise, and each of these assumptions, when it fails, restores
the dependence on the drive. A finite optical depth mixes propagation
effects into the observable. In a hot vapor, velocity averaging combines
correlations from atomic velocity classes at different one-photon
detunings, a complication that the four-wave-mixing work of Section 8.8
eventually has to confront. The laser's residual intensity noise, absent
from the model, is a common-mode input that adds a positive background
independent of the medium. Zheng and co-workers showed that including it
in the treatment measurably broadens the correlation resonance with
increasing power in a buffer-gas cell \cite{r21}.

\emph{Stray fields and geometry.} The universal scaling assumes that the
applied longitudinal field is the only source of two-photon detuning.
Residual transverse fields, field gradients and imperfect shielding all
add an uncontrolled contribution to both \(\Delta\) and \(\gamma_{bc}\),
broadening the curve and biasing any quantity extracted from it; the
transit of atoms through the beam has the same effect. These are
ordinary reasons why a room-temperature measurement departs from a
thin-medium calculation, and they are probably the origin of the
asymmetry in the wings noted at the end of Section 4.2.

Finally, \emph{the range demonstrated experimentally is narrower than
the claim suggests.} The cleanest demonstration to date found the width
of the correlation peak essentially constant as the intensity was
raised, in cold atoms where Doppler and transit effects are absent
\cite{r4}. It covered only a factor of about 10--17 in optical power
(30--300 μW for rubidium and 30--500 μW for cesium), in two atomic
species. This is a strong result, but it is not the same as an
insensitivity established across the full range of powers, optical
depths and cell types in which the technique is used. Neither edge of
that range has been probed.

None of this invalidates Section 4.2. The switching curve is stable over
a wide operating range, wide enough for a comparison between theory and
experiment with one scale factor to work well. But this stability is an
approximation with identifiable failure modes, and any metrological use
of the curve, including that of Section 7, must first establish that it
operates well within them.

\subsection{Frequency-domain and perturbative treatments}\label{sec:4.4}

A complementary line of theory works with the frequency-resolved
correlation \(C(\omega)\) and treats the phase noise perturbatively.
Flórez, González and Martinelli developed a perturbative
frequency-domain approach to the frequency-resolved intensity
correlation C(ω) under EIT conditions \cite{r32}. The motivation, stated
in that work and in the thesis behind it, was analytic insight. Noise
correlation spectra had been well described by numerical calculations,
including the group's own frequency-domain ones, but without an
analytical model it was hard to see which atomic properties set the
spectroscopic features \cite{r32,r43}. Expanding perturbatively in the
noise strength sacrifices the ability to treat a strongly
phase-diffusing drive, but gains analytic transparency in the regime
where the expansion is valid.

Because it is non-perturbative in the noise, the stochastic treatment
remains valid for a free-running diode laser with a broad linewidth, but
it delivers its results as simulated time series. The perturbative
frequency-domain treatment gives lineshapes and widths analytically, at
the cost of assuming that the noise is weak enough for a low-order
expansion. Whether, and where, these two regimes overlap has not been
determined.

\subsection{Master-equation and Lindblad approaches}\label{sec:4.5}

A third approach describes the three-level system with a Lindblad master
equation, the standard equation of motion for the density matrix of a
system coupled to its environment, and computes two-time
intensity--intensity correlations directly from the quantum regression
theorem. DiLoreto and Rangan used this method to analyze three-level
systems driven by two detuned fields and found low-frequency quantum
beats in the intensity--intensity correlations \cite{r33,r44}. The
framework handles structured dissipation easily but treats the drive as
monochromatic, the reverse of the stochastic treatment, which handles a
noisy drive but keeps the dissipation simple. The two approaches are
therefore complementary. One pairs a realistic laser with a simple
environment, the other an idealized laser with a structured one. No
published work yet combines a noisy drive with structured dissipation.

\subsection{Summary of frameworks}\label{sec:4.6}

Table 1 compares the three families of approaches.

{\small
\begin{longtable}{@{}
  >{\raggedright\arraybackslash}p{(\columnwidth - 6\tabcolsep) * \real{0.2158}}
  >{\raggedright\arraybackslash}p{(\columnwidth - 6\tabcolsep) * \real{0.3079}}
  >{\raggedright\arraybackslash}p{(\columnwidth - 6\tabcolsep) * \real{0.2285}}
  >{\raggedright\arraybackslash}p{(\columnwidth - 6\tabcolsep) * \real{0.2478}}@{}}
\caption{Theoretical frameworks applied to intensity--intensity noise correlations in coherently prepared media, and what each is suited to calculate.}\label{tab:1}\\
\toprule\noalign{}
\begin{minipage}[b]{\linewidth}\raggedright
\textbf{Framework}
\end{minipage} & \begin{minipage}[b]{\linewidth}\raggedright
\textbf{Drive}
\end{minipage} & \begin{minipage}[b]{\linewidth}\raggedright
\textbf{Delivers}
\end{minipage} & \begin{minipage}[b]{\linewidth}\raggedright
\textbf{Limitation}
\end{minipage} \\
\midrule\noalign{}
\endhead
\bottomrule\noalign{}
\endlastfoot
Stochastic density matrix \cite{r16} & finite-bandwidth, multiplicative
phase noise (non-perturbative) & \(G^{(2)}(\tau)\) time series;
universal scaling in \(\gamma_{bc}\) & results are numerical; no closed
form for the linewidth \\
Perturbative frequency domain \cite{r32} & weak phase noise &
\(C(\omega)\) lineshapes analytically & restricted to weak noise \\
Lindblad master equation \cite{r33} & monochromatic (idealized) & two-time
correlations, quantum beats & no laser noise; cannot produce the
switching curve \\
\end{longtable}
}

\section{Experimental methods and quantitative benchmarks}\label{sec:5}

\subsection{The standard apparatus}\label{sec:5.1}

The optical arrangement of the original measurement \cite{r14} has been
reused, with variations, in much of the subsequent literature (Figure
3).

An external-cavity diode laser is tuned to the \textsuperscript{87}Rb
D\textsubscript{1} line,
\(5S_{1/2}(F = 2) \leftrightarrow 5P_{1/2}(F' = 1)\). The beam is split,
and one arm passes through a \(\lambda/2\) plate so that the two arms
carry orthogonal linear polarizations. The arms are recombined on a
polarizing beam splitter, and a \(\lambda/4\) plate converts the pair
into two counter-rotating circular polarizations. After the cell, a
second \(\lambda/4\) plate and polarizing beam splitter direct the two
circular components onto two fast photodiodes with matched optical path
lengths. The 7.5 cm cell contains rubidium at natural abundance,
\(N \sim 10^{12}\) cm\textsuperscript{-3}, and sits inside a two-layer
magnetic shield. The detection band spans 75 kHz--1.2 GHz; the signal is
calibrated as Signal {[}V{]} \(= 500\) {[}V/W{]} \(\times I_{laser}\);
records last 10 μs; and the total optical power of 1 mW is divided into
two 0.5 mW beams of 0.1 cm diameter. A longitudinal magnetic field sets
the two-photon detuning, \(\Delta = aB\), and is the single parameter
that tunes the system from correlation to anticorrelation.

Four features of this design matter when results are compared across
papers.

\emph{A single laser, split into two beams.} Deriving both modes from
one source is what makes the phase noise common. If two independent
diode lasers are used instead, as Cruz and co-workers did on the
\textsuperscript{85}Rb D\textsubscript{2} line in a room-temperature
cell \cite{r45}, the problem changes fundamentally. The two fields then
carry uncorrelated phase noise, and any correlation between them is
created by the medium alone. Cruz and co-workers nevertheless observed
correlations as large as 0.65 that turned into anticorrelations of
similar size as the optical intensity was raised, and modelled them with
independent phase-diffusion processes, including the Doppler width
\cite{r45}. Optical intensity is thus a second control parameter for the
sign, alongside the magnetic field. Neither configuration is wrong, but
they are not interchangeable, and comparisons across the literature must
keep track of which was used. Xiao and co-workers examined the
single-laser configuration specifically, using both Zeeman and hyperfine
EIT. They showed that whether laser phase noise appears in the
transmitted spectrum as a suppressed dip or an enhanced peak depends on
the relative size of the one-photon and two-photon detuning noise
\cite{r46}.

\begin{figure}[!htbp]
\centering
\includegraphics[width=0.92\linewidth]{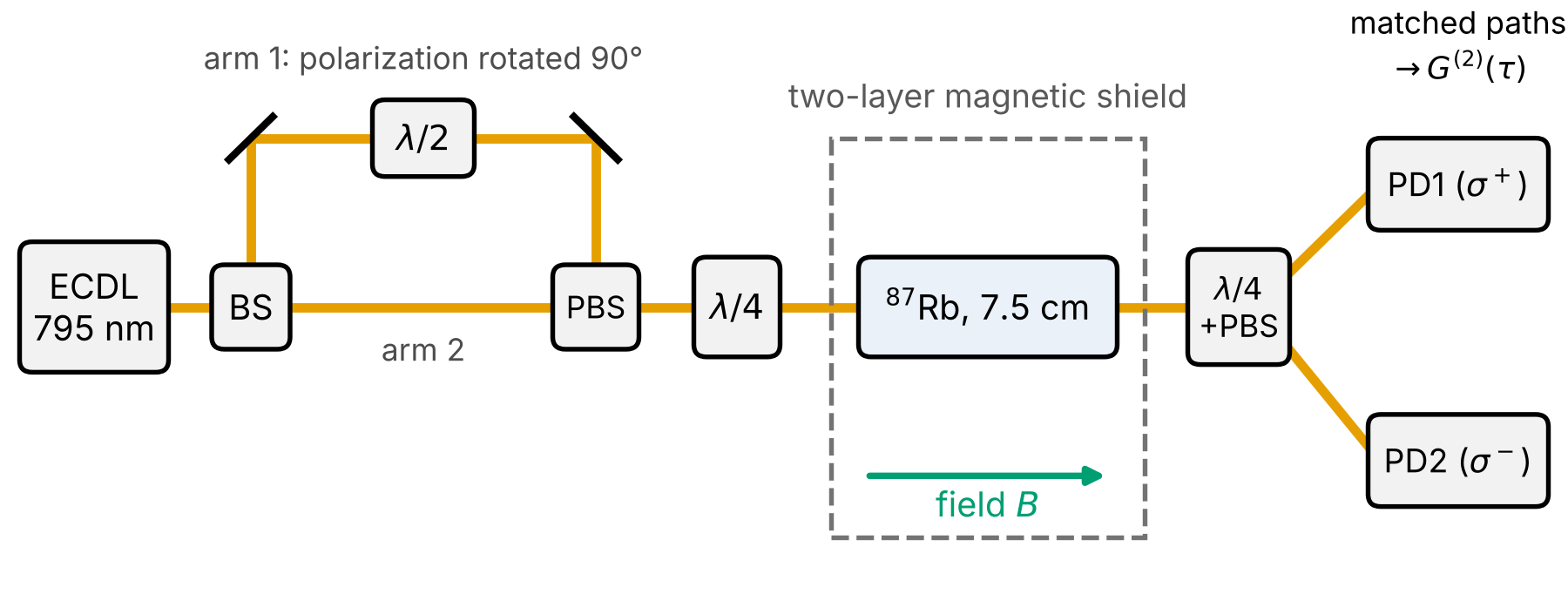}
\caption{Standard apparatus for two-mode intensity-noise correlation measurements in a coherently prepared vapor, after Ref.~\cite{r14} (schematic). The beam of a single external-cavity diode laser is divided by a beam splitter (BS) into two arms. A half-wave plate (λ/2) in one arm rotates its polarization by 90°, and the two arms are recombined on a polarizing beam splitter (PBS) with orthogonal linear polarizations. A quarter-wave plate (λ/4) then converts them into two counter-rotating circular components. After the shielded Rb cell, a second quarter-wave plate and PBS separate the two components onto photodiodes with matched path lengths. The longitudinal field B is the single control parameter.}\label{fig:3}
\end{figure}

\emph{Matched detection path lengths.} A correlation measured at
\(\tau = 0\) should reflect the medium, not a relative optical delay. A
path mismatch of a few centimeters corresponds to about 100 ps, which is
short compared with both the atomic timescales and the 1.2 GHz detection
band. Small mismatches are harmless. The requirement matters only when
the delay approaches the inverse detection bandwidth, at which point the
peak of \(G^{(2)}(\tau)\) shifts away from \(\tau = 0\).

\emph{Magnetic shielding.} The applied longitudinal field must be the
only source of two-photon detuning. Residual transverse fields and
gradients add an uncontrolled contribution to \(\Delta\) and
\(\gamma_{bc}\), broadening the switching curve and biasing any
decoherence rate extracted from it; this is the requirement on which the
protocol of Section 7 depends most. Because the protocol has not yet
been applied independently (Section 7.2), there are no reproductions to
show how often imperfect shielding is the limiting factor. The general
point remains that any residual field passing through the shield is
carried directly into the extracted rate.

\emph{Detection bandwidth.} The measured \(G^{(2)}(0)\) integrates over
the detection band. The noise generated by the medium lies at
frequencies comparable to the relevant relaxation rates. If the
detection band does not reach low enough frequencies, or if the record
is too short to resolve the ground-state timescale, the observed
contrast is reduced. The measured signal gives no sign that this has
happened. The background levels quoted alongside the correlation values
in Section 5.2 largely reflect this effect.

\subsection{Quantitative benchmarks}\label{sec:5.2}

Table 2 collects the values from the original measurements, which later
work can adopt, reproduce or contradict, and to which the rest of the
review refers.

{\small
\begin{longtable}{@{}
  >{\raggedright\arraybackslash}p{(\columnwidth - 2\tabcolsep) * \real{0.6512}}
  >{\raggedright\arraybackslash}p{(\columnwidth - 2\tabcolsep) * \real{0.3488}}@{}}
\caption{Quantitative benchmarks from the original correlation-switching measurement and its stochastic treatment \cite{r14,r16}. For reference, the same sample showed a peak EIT transmission of 0.63 compared with a linear transmission below 1\%, an EIT FWHM of 0.85 G at 0.5 mW/beam compared with a 0.16 G low-power limit, and an EIT spectral width of 1.2 MHz.}\label{tab:2}\\
\toprule\noalign{}
\begin{minipage}[b]{\linewidth}\raggedright
\textbf{Quantity}
\end{minipage} & \begin{minipage}[b]{\linewidth}\raggedright
\textbf{Value}
\end{minipage} \\
\midrule\noalign{}
\endhead
\bottomrule\noalign{}
\endlastfoot
\(G^{(2)}(0)\), \(B = 0\), overlapped beams & \(+ 0.9\) (background
\(\approx 0.15\)) \\
\(G^{(2)}(0)\), anticorrelated (representative field \(B = - 0.47\) G) &
\(- 0.9\) (anticorrelation dip) \\
\(G^{(2)}(0)\), beams separated by 0.3 cm (diameter 0.1 cm) & \(+ 0.7\)
(background \(\approx 0.3\)) \\
Width of the correlation-versus-\(B\) resonance, 0.5 mW/beam & 0.24 G \\
EIT transmission width, 0.5 mW/beam & 0.85 G \\
Ratio of correlation resonance to EIT width, same power &
\(\approx 4 \times\) narrower \\
Extracted ground-state decoherence rate \(\gamma_{bc}\) & \(\approx 1\)
MHz (scale factor \(\alpha \approx 1\) G) \\
Zeeman conversion constant \(a\) & 1.4 MHz/G \\
\end{longtable}
}

Rows four and six give the practical case for the method. On the same
sample and at the same power, the correlation resonance is about four
times narrower than the EIT transmission resonance. A narrower line is
not automatically a resolved one, however; Section 6.4 describes the
separate measurement in which this difference in width actually
separated peaks that overlap in transmission. Row three is the
spatial-separation control discussed above. Row seven is the outcome of
the protocol of Section 7: one scale factor matched to a universal
curve, with nothing else adjusted. It therefore carries the systematic
errors of the field calibration, the shielding and the thin-medium
approximation (Section 4.3).

The author is not aware of any later measurement that has reproduced or
contradicted these values on a different sample. Every entry in Table 2
still rests on a single experiment in a single cell. Later papers cite
these numbers without checking them, and the original measurement still
stands alone. The closest check is indirect: in cold
\textsuperscript{87}Rb, de Almeida and co-workers measured zero-delay
correlations above 0.95 between two transmitted beams and described
near-perfect correlation as a known result of Ref.~\cite{r16} \cite{r27}.
That concerns only the positive branch, without a magnetic-field scan.
The switching values themselves remain unreproduced. Normalized,
dimensionless quantities are exactly the kind of result that should be
transferable between laboratories, but whether these particular values
are has never been tested.

\subsection{Variants of the apparatus}\label{sec:5.3}

Most of the experimental literature falls into four families of
variations on this design.

\emph{Cold atoms.} Replacing the vapor cell with a magneto-optical trap
removes both Doppler broadening and transit-time relaxation, leaving
stray fields and the trapping light as the main sources of residual
ground-state decoherence. Felinto and co-workers, in a collaboration
between groups in Recife (UFPE) and São Paulo (USP), performed the
measurement in a cold ensemble from a magneto-optical trap \cite{r19}. The
same collaboration later extended it to cold \textsuperscript{133}Cs and
\textsuperscript{85}Rb \cite{r4} and to cold \textsuperscript{87}Rb on the
\(5S_{1/2}(F = 1) \rightarrow 5P_{3/2}(F' = 1)\) transition \cite{r5}. The
approximations of Section 3 hold best here, with no Doppler averaging
and no transit-time relaxation, and where the power-broadening-free
measurement of Section 6.2 was made.

\emph{Buffer gases.} Adding neon, or neon and argon, to a warm cell has
two simultaneous effects: diffusion confines the atoms and suppresses
transit-time relaxation, while collisions introduce dephasing. The
resulting \(\gamma_{bc}\) falls in a range of direct practical interest
for chip-scale devices. Zheng, Green, Crescimanno and O'Leary studied
Zeeman EIT in the Hanle configuration on the \textsuperscript{87}Rb
D\textsubscript{1} line, in warm cells containing either 2 Torr of neon
or 10 Torr of neon and argon (5 Torr each) \cite{r21}. Whitenack and
co-workers used a warm (44 °C) isotopically enriched
\textsuperscript{87}Rb cell with the same buffer gases, driven by a
free-running 795 nm diode laser on the D\textsubscript{1} line. They
added a third polarization-detection channel to recover the full
intensity-noise correlation content rather than only the noise on one
fixed pair of polarization modes \cite{r22}.

\emph{Isotopically enriched cells and alternative level schemes.}
Enriched \textsuperscript{87}Rb removes absorption by the other isotope
and simplifies the level structure. Feng and colleagues used such cells
to measure the linewidth of the transition from correlation to
anticorrelation \cite{r17}, and Li, Feng and Xiao used magnetically split
multipeak resonances in the same spirit \cite{r18}.

\emph{Cavities.} Placing the medium inside a resonator changes the mode
structure available to the correlation. Rao and co-workers showed that
for \emph{N} cold Λ-type atoms held in one arm of a two-mode loop
cavity, the cross-intensity correlation reverses sign, from negative to
positive, under a symmetric variation of the two atom--pump detunings.
The switch is thus driven by detuning, not by a magnetic field. They
explained the mechanism with a dressed-state picture and Bogoliubov
modes \cite{r23}. Zhou and colleagues examined a related phase correlation
that builds up during two-photon resonance inside an active cavity
containing a gain medium \cite{r47}.

\section{Sub-natural-linewidth and power-broadening-free correlation spectroscopy}\label{sec:6}

Through \(a = 1.4\) MHz/G, the 0.24 G correlation resonance of Table 2
corresponds to about 0.34 MHz. This is roughly four times below the EIT
transmission width on the same sample at the same power, and about
seventeen times below the 5.75 MHz natural linewidth of the rubidium
D\textsubscript{1} transition. The second comparison is not strictly
like with like, since it sets a two-photon feature against a one-photon
linewidth, but it does establish that the correlation resonance is not
limited by the lifetime of the excited state.

\subsection{Why the correlation resonance is narrow}\label{sec:6.1}

Conventional EIT spectroscopy measures transmission, and in the standard
treatment the half-width of the transmission window is

\[\Gamma_{EIT} \sim \gamma_{bc} + \frac{|\Omega|^{2}}{\gamma},\]

so at any drive strong enough to give a useful signal, the second term
dominates and the measured width reflects the drive more than the
sample. This is power broadening. For this reason, an EIT linewidth
quoted without its optical power cannot be meaningfully compared with
another.

The correlation observable avoids this. Because the sign change is a
ground-state effect (Section 3.4), the sign of \(G^{(2)}(0)\) is
governed by the competition between the two-photon detuning \(\Delta\)
and the ground-state decoherence rate \(\gamma_{bc}\). The position of
the zero crossing is therefore set by the comparison
\(\Delta \sim \gamma_{bc}\) rather than mainly by the drive. Computed
exactly, the crossing in Figure 2 lies at 0.074--0.078 of the
decoherence rate across the decade of rates in Figure 2(a), and at
0.064--0.094 of it across the drives 0.6γ--4γ in Figure 2(b).

The stochastic treatment makes this quantitative, as described in
Section 4.2: the computed switching curves for
\(\Omega_{1} = 0.6,0.8,\ldots,4\,\gamma\) nearly coincide, so above
\(\Omega_{1} \gtrsim \gamma\) the correlation resonance does not broaden
with the drive, while its width remains proportional to \(\gamma_{bc}\)
\cite{r16}. The room-temperature experiment shows what this means in
practice, for one sample at one power: 0.24 G compared with 0.85 G, a
factor of about four \cite{r14}. Figure 2(c,d) shows this width comparison
computed from the same model: the EIT window broadens steadily as the
drive increases, whereas the correlation peak does not. In this
idealized Doppler-free model the difference is far larger than the
factor of four measured in a warm vapor.

\subsection{Power-broadening-free correlation spectroscopy}\label{sec:6.2}

The power-broadening-free behavior of the correlation resonance was
first hinted at theoretically by the author and his collaborators. The
drive-independence of the switching curve (Section 4.2) is this
property, and it appears in the 2006 preprint of Ref.~\cite{r16}, four
years ahead of its journal version and before the effect had been
measured systematically. In 2013, Flórez and co-workers demonstrated
power-broadening-free correlation spectroscopy in cold atoms. Scanning
the excitation power over roughly a decade in \textsuperscript{133}Cs
and \textsuperscript{85}Rb magneto-optical traps, they found that the
linewidth of the EIT correlation peak, as defined in their work,
remained essentially unchanged while the transmission peak broadened
steadily \cite{r4}. This was a systematic cold-atom demonstration, not the
first observation of the effect. As Ref.~\cite{r4} notes, the
insensitivity of the correlation-peak width to the field intensity had
already been observed in vapor cells \cite{r14,r46}. In Ref.~\cite{r4}, the
quantities found to be insensitive to power are the linewidth of the EIT
correlation peak and the transition point between correlation and
anticorrelation. These are related to, but not automatically the same
as, the zero crossing in magnetic field discussed in Section 6.1 or the
transition linewidth of Ref.~\cite{r17} (Section 6.3), and comparisons
between them should keep the definitions distinct. Within the range
tested, at least, the power can be increased without loss of resolution.

The phrase ``free of power broadening'' should be read with the limits
of Section 4.3 in mind. Experimentally, it has been established over
about one decade of optical power in a magneto-optical trap. Below the
drive threshold the insensitivity fails, and the weak-drive curves in
the computed family are broader by a large factor. A measurement made at
low power to avoid disturbing the sample therefore lies in exactly the
regime where the width depends on power. At high power, optical pumping,
light shifts and saturation should eventually restore a power dependence
that no experiment has yet looked for. In a warm cell, moreover, the
laser's own residual intensity noise can broaden the correlation
resonance with power in ways the phase-noise-only treatment does not
predict, as Zheng and co-workers found in their buffer-gas measurements
\cite{r21}. Where the plateau begins and ends has not been documented
(Section 11).

\subsection{The transition linewidth as a distinct observable}\label{sec:6.3}

If the correlation curve crosses zero, the \emph{width of that
crossing}, that is, the range of detuning over which the correlation
passes from positive to negative, is a spectroscopic quantity in its own
right. No transmission measurement provides it. Feng, Li, Zhang, Wang
and Xiao introduced and measured this quantity under the name transition
linewidth of the cross-correlation \cite{r17}. They found that it can lie
below the (effective) coherence-lifetime limit of the EIT width and is
limited by competing noise sources, and their numerical model shows that
it depends on the laser linewidth and the laser power \cite{r17}.

This quantity should not be confused with the resonance width of Table
2. The resonance width is the full width of the correlation-versus-field
feature, whereas the transition linewidth describes how sharp the sign
change is at its center. In the idealized model of Section 4.2 both are
set by \(\gamma_{bc}\) but weight the underlying dynamics differently.
Measuring both would give two independent values of the ground-state
rate and test the model at the same time. The dependence of the
transition linewidth on laser linewidth and power found in Ref.~\cite{r17}
shows that such a test would also have to account for the laser. This
has not yet been done; the papers that introduced the two quantities do
not compare them.

\subsection{Resolving structure that transmission cannot}\label{sec:6.4}

Li, Feng and Xiao applied the technique to hyperfine EIT in
\textsuperscript{87}Rb with magnetically split multipeak resonances, and
showed that a cross-correlation resonance narrower than the transit-time
linewidth separates individual peaks that overlap in the transmission
spectrum \cite{r18}. The demonstration is specific to magnetically split
hyperfine peaks. It has not been tested on other spectra, whose
structure may not appear in the correlation signal.

\subsection{Sideband resonances and the frequency-domain picture}\label{sec:6.5}

Working in the frequency domain, Flórez, Kumar, Theophilo, Nussenzveig
and Martinelli found light sideband resonances in the frequency-resolved
correlation C(ω) of cold \textsuperscript{87}Rb under EIT conditions,
using two orthogonally circularly polarized fields coupling Zeeman
sublevels. The polarization geometry is the same as in the original
room-temperature experiment, now applied to trapped atoms \cite{r5}. The
resonances appear only at suitable analysis frequencies, and the authors
show that the sidebands cannot be resolved in the correlation
coefficient unless the one-photon detuning is nonzero. This shows that
the two descriptions are not redundant.

\subsection{Cold atoms and mechanical effects}\label{sec:6.6}

Theophilo and co-workers applied the technique to the light forces
themselves, using noise correlation spectroscopy to probe the mechanical
effect of light on cold atoms \cite{r6}. The correlation signal thus also
responds to atomic motion, which has little to do with the
internal-state spectroscopy for which the method was developed. The
observable responds to any process that changes the two coherences
differently, and this need not be an internal-state process. In this
case, it was the light force. The same forces produce a spectral
asymmetry in cold-atom noise correlation spectra, which the authors
describe as a hysteresis effect \cite{r6,r43}. Asymmetry in a measured
switching curve therefore has more than one possible origin besides the
stray fields discussed in Section 4.3.

\section{Decoherence metrology from the switching curve}\label{sec:7}

\subsection{The protocol}\label{sec:7.1}

Section 4.2 showed that the correlation-switching curve has a universal
shape whose only scale is \(\gamma_{bc}\), and the author and his
co-workers proposed a measurement protocol that follows almost directly
from this fact \cite{r16}. The experimental control parameter is the
longitudinal field \(B\), which sets the two-photon detuning through
\(\Delta = aB\), with \(a = 1.4\) MHz/G for the \textsuperscript{87}Rb
Zeeman pair addressed. Writing
\(B = \alpha\left( \Delta/\gamma_{bc} \right)\) and matching the
computed universal curve to the measured \(G^{(2)}(0,B)\) curve with the
single scale factor \(\alpha\) gives

\[\gamma_{bc} = \alpha a.\]

For the rubidium vapor of the original experiment this yields
\(\alpha \approx 1\) G, and therefore \(\gamma_{bc} \approx 1\) MHz.
This is an order-of-magnitude value, since the curve match fixes α only
to about one Gauss, while the conversion involves the 1.4 MHz/G Zeeman
constant. No other adjustable quantity enters the comparison: neither
the population decay rates nor the drive strength.

The protocol has three features of interest for metrology. First, it is
calibration-free in a specific and narrow sense: it requires no
reference sample, no independently known relaxation rate and no absolute
intensity calibration. The reason is that the observable is normalized
and the conversion from field to detuning is a known atomic constant.
One calibration remains. The field axis must be known in Gauss. Errors
in the solenoid calibration and the residual field inside the shield
therefore propagate directly into \(\gamma_{bc}\). Second, it is
independent of the population decay rates, which do not enter the family
of curves computed in Section 4.2. Third, it does not broaden with drive
strength over the plateau computed there (it narrows slightly and levels
off). It therefore avoids the weak-drive limit that a measurement of
\(\gamma_{bc}\) from the transmission linewidth would need in order to
escape power broadening.

The cautions of Section 4.3 bear directly on how far any number obtained
with this protocol can be trusted. The extraction assumes that the
measurement lies on the drive plateau, above the threshold. It also
assumes that the applied longitudinal field is the only source of
two-photon detuning, so that residual transverse fields, gradients and
transit-time relaxation do not add an unmeasured contribution to
\(\gamma_{bc}\). Finally, it assumes that the medium is optically thin
and free of velocity averaging, with negligible laser amplitude noise.
In the room-temperature comparison, the agreement is close through the
peak, the zero crossing and the approach to anticorrelation, but the
measured points depart from the symmetric computed curve in the wings
and are visibly asymmetric about \(B = 0\). The result is therefore a
good estimate obtained with no free parameters beyond the single scale
factor, not a high-accuracy determination with a small stated
uncertainty. Assigning a real uncertainty to \(\gamma_{bc}\) with this
protocol would require a systematic study of all the factors listed
above, and no such study exists.

\subsection{Status}\label{sec:7.2}

Felinto and co-workers studied the same noise correlation spectra in a
cold ensemble, writing the correlation in terms of the absorptive parts
of the atomic polarizations \cite{r19}. They interpret the
power-independent linewidth differently, however: as the result of the
steep normal dispersion near two-photon resonance, which laser phase
noise maps into fluctuations of the absorption and which competes with
the transparency itself. This is a complementary physical picture of the
same spectra rather than an application of the field-scale protocol, and
the author has found no independent application of the protocol. Hétet
also determined the ground-state decoherence rate of a buffer-gas
system, but from the efficiency of EIT light storage compared against a
numerical model rather than from a switching field \cite{r48}. The
decoherence rate is of general interest and can be obtained in several
ways; what is specific to this protocol is that it reads the rate
directly from the correlation switching, with one adjusted scale factor
and nothing else.

\(\gamma_{bc}\) is the key figure of merit for CPT clocks \cite{r13},
EIT-based magnetometers \cite{r12} and ground-state quantum memories
\cite{r11}, and it is the parameter that cell manufacturers tune when they
choose buffer gases and wall coatings. A measurement that requires only
two photodiodes, a shielded solenoid and a single adjusted scale factor
should therefore be attractive. Why it has not been adopted is not
discussed anywhere in the literature. One possible explanation is
geometric: the protocol, as formulated, requires the specific Λ
arrangement with two circular components. Groups working with buffer-gas
cells have mostly used Hanle configurations, in which the mapping from
field to two-photon detuning is different. This is a conjecture. No
published source gives a reason for setting the method aside, and a
search of the buffer-gas papers for one finds none.

\section{Extensions and new effects}\label{sec:8}

In the two decades since the original observation, the technique has
been carried into a wide range of systems. The following subsections are
organized by the physics that changes.

\subsection{From induced transparency to induced absorption}\label{sec:8.1}

Electromagnetically induced absorption (EIA) is the counterpart of EIT
with the opposite sign. It occurs on degenerate transitions with
\(F' = F + 1\), where transfer of coherence from the excited state
enhances rather than suppresses absorption \cite{r49,r50}. Lee and Moon
measured the intensity fluctuations under EIA on the
\(5S_{1/2}(F = 2)\)--\(5P_{3/2}(F' = 3)\) transition of
\textsuperscript{87}Rb and found that the two circularly polarized
components switch between correlation and anticorrelation. They
attributed this to atomic coherence spontaneously transferred within the
EIA medium \cite{r20}. This is a different route to the same sign
structure, in a system where the resonance is inverted.

EIT and EIA differ in whether the ground-state coherence is dark or
bright with respect to the driving field. In principle, the correlation
observable should distinguish between them on a given degenerate
transition. This would be useful, because absorption spectroscopy alone
can leave the distinction ambiguous when several \(F \rightarrow F'\)
components overlap. Ref.~\cite{r20} establishes that the sign structure
survives under EIA, but using it to tell EIT from EIA in practice is a
proposal made here, not a reported result.

\subsection{Polarization structure and the propagation eigenbasis}\label{sec:8.2}

Lee, Jeong and Moon studied Hanle EIT on \textsuperscript{87}Rb
\(5S_{1/2}(F = 2) \rightarrow 5P_{1/2}(F' = 1)\) in the split-beam
geometry with orthogonal polarizations, and showed that selecting a
polarization basis after the cell (postselection) changes what the
intensity correlation reports \cite{r51}. The reason is that the medium in
a longitudinal field is birefringent and dichroic. The polarization
basis in which the two propagating modes are independent (the
propagation eigenbasis) generally differs from the basis in which the
beams were prepared. Whitenack and co-workers took this further, noting
that earlier EIT noise spectroscopy had been confined to the
\(\sigma^{\pm}\) propagation eigenbasis, and extending the analysis
beyond it \cite{r22}.

The correlation between two detected channels therefore depends on the
relationship among three bases: the preparation basis, the eigenbasis of
the medium and the detection basis. Polarization thus becomes a second
control parameter alongside the magnetic field, and a correlation value
quoted without its analysis basis is, strictly speaking, incomplete.

\subsection{Collisional environment and the laser's own intensity noise}\label{sec:8.3}

Buffer gases have two opposing effects: they lengthen the ground-state
coherence time by confining atoms diffusively within the beam, and
shorten it through collisional dephasing. The net \(\gamma_{bc}\)
therefore varies non-monotonically with buffer-gas pressure, which is
why the choice of buffer gas has become a subject of its own in the
development of chip-scale clocks and magnetometers.

Zheng, Green, Crescimanno and O'Leary examined Zeeman EIT on the
\textsuperscript{87}Rb D\textsubscript{1} line in warm neon and
neon--argon cells. They extended the theory by including the laser's
residual intensity noise as a second noise channel alongside the phase
noise \cite{r21}. They found that this additional term affects the weaker
of their two EIT systems most strongly, causing its noise correlation to
broaden with power more than the phase-noise-only treatment would
predict.

It is the only extension of the stochastic model that the author has
found in the literature. It shows that the assumption of negligible
diode-laser amplitude noise is good but not exact, and that where it
fails, the correlation acquires a power dependence that the
phase-noise-only treatment does not contain.

\subsection{From two fields to many: multiwave mixing}\label{sec:8.4}

The definition of the observable applies to any two fields whose
intensities can be measured. A natural next step is to study media that
carry more than two fields at once.

Yang, Sheng, Khadka and Xiao increased the number of mutually correlated
modes to four and then six, using triple- and quadruple-Λ level schemes
in \textsuperscript{85}Rb on the D\textsubscript{1} line in a
neon-buffered cell (3 Torr). The prepared spin coherence links the
generated and applied fields \cite{r25}. Zheng, Khadka, Song, Zhang and
Xiao studied noise correlations among three fields produced by a
third-order nonlinear mixing process \cite{r24}. Sheng, Wu, Yang, Khadka
and Xiao mapped the pairwise noise correlations inside a doubly resonant
atomic optical parametric oscillator \cite{r52}.

With more than two fields there is more than one pairwise correlation,
and the pattern of which pairs are correlated and which anticorrelated
becomes an observable in its own right. Lan and co-workers showed that
these correlations can be controlled by dressing, that is, by adding a
field that suppresses or enhances particular channels \cite{r53}. Li and
colleagues showed that the polarization configuration itself controls
the noise correlation of four-wave-mixing processes in rubidium vapor
\cite{r54}.

\subsection{Rydberg-dressed systems}\label{sec:8.5}

Zhang and co-workers observed intensity-noise correlations among
six-wave-mixing, fluorescence and probe-transmission signals in a
\emph{K}-type five-level thermal \textsuperscript{85}Rb ensemble coupled
to Rydberg \emph{n}D\textsubscript{5/2} states with \emph{n} = 37, 45,
54 and 63. They controlled the correlation through Rydberg dressing
\cite{r26}. Dressing introduces a controllable interaction between atoms.
The coherences being read out are no longer those of independent atoms,
which makes this a notable configuration. The paper uses the dressing to
control the correlation. A measurement of interaction-induced dephasing
from the correlation, which this configuration appears to allow, has not
yet been reported.

\subsection{Four-wave mixing: correlations between generated signals}\label{sec:8.6}

The most active recent line of work extends the observable to forward
four-wave mixing, in which the two correlated modes are no longer two
transmitted beams but two \emph{generated} fields. (Correlations between
generated and applied fields appear in the multiwave-mixing, Rydberg and
vortex work \cite{r25,r26,r31}.)

de Almeida, da Motta and Vianna sent two input beams derived from a
single continuous-wave laser through cold \textsuperscript{87}Rb in a
magneto-optical trap, on the D\textsubscript{2} transition F = 2 → F′ =
3. They measured the intensity correlations between the two forward
four-wave-mixing signals they generate, as well as between the two
transmitted beams \cite{r27}. They used the stochastic model of Section
4.1 to explain the transmission correlations; the four-wave-mixing
signals were modelled only in their later work \cite{r28}. The two
polarization configurations access entirely different level structures:
with linear, orthogonal polarizations the light interacts with a sum of
\(\Lambda\) systems formed by the Zeeman sublevels. With circular,
parallel polarizations the system reduces to a pure two-level atom. A
single set of measurements thus compares two physically distinct routes
to the correlation. The experiments record about 100 μs of signal at
roughly 1 ns resolution, and the theory uses Itô calculus with a
stochastic Runge--Kutta solver rather than the Euler--Maruyama scheme of
Section 4.1 \cite{r27,r28}. Miranda, Rosas, Passerat de Silans and Lopez
developed a companion theory for two symmetric four-wave-mixing signals
in a two-level atom, with Wiener phase noise (an Ornstein--Uhlenbeck
process gave similar results). It predicts Rabi oscillations in the
correlation function and, at weak detuning, a crossover from correlation
to anticorrelation \cite{r55}.

The two-level results show that the Λ configuration is not required. The
correlation switching of Section 3 relies on the antisymmetric
ground-state population fluctuation of a Λ system, a mechanism that does
not exist in a two-level atom. Yet strong correlations between the two
generated signals, and between the transmitted beams, are observed there
as well \cite{r27,r28}, and the two-level model of Miranda et al. even
predicts a change of sign \cite{r55}. In these works the correlations are
attributed to the same resonant conversion of phase noise into amplitude
noise, read out in the correlation function at the generalized Rabi
frequency. The Λ mechanism is therefore sufficient for sign switching
but not necessary. What is still missing is a physical account of the
two-level sign change as explicit as the one in Section 3.4, and this is
one of the clearest open theoretical questions in the field.

\subsection{Rabi oscillations in the correlation signal}\label{sec:8.7}

In the same cold-atom experiments, de Almeida, da Motta and Vianna found
that the intensity correlations, both between the two four-wave-mixing
signals and between the transmitted beams, oscillate at the Rabi
frequency. They reproduced the transmission results with time series
generated from the stochastic optical Bloch equations \cite{r27,r56}. The
correlation function acquires coherent oscillatory structure at the
generalized Rabi frequency, so \(G^{(2)}(\tau)\) now carries dynamical
information about the driven system, not only the static information
contained in \(G^{(2)}(0)\).

This result also connects to the sideband resonances found in the
frequency domain \cite{r5}. Oscillations at the Rabi frequency in
\(G^{(2)}(\tau)\) and sidebands at the Rabi frequency in \(C(\omega)\)
are the same physics viewed through a Fourier transform, which is one
more reason to reconcile the two descriptions.

\subsection{The Doppler-broadened extension}\label{sec:8.8}

de Almeida, Alvarez, de Melo, da Motta, de Souza and Vianna took the
four-wave-mixing correlation measurement from the cold-atom sample to a
hot \textsuperscript{85}Rb vapor on the D\textsubscript{2} line, driven
by a single continuous-wave laser. For circular, parallel input
polarizations, the forward geometry forms a pure two-level
configuration. They adopted the stochastic model and extended it, with
separate responses for the four-wave-mixing and transmitted signals
integrated over the atomic velocity distribution \cite{r28}. They report
an oscillatory dependence of the cross-correlations on incident power,
relate it to Rabi oscillations, and show how the separate velocity
classes shape the observed response.

This step treats explicitly the approximation noted in Section 3.2; Cruz
et al. had already explored the role of the Doppler width numerically
for two-laser EIT correlations \cite{r45}. The original derivation
neglected Doppler averaging because the two-photon detuning of a
co-propagating Λ system is Doppler-free to first order. That argument
does not hold for a four-wave-mixing geometry in a hot vapor, where the
velocity average must be carried out explicitly. Different velocity
classes lie at different one-photon detunings, convert phase noise into
intensity noise with different efficiencies, and contribute correlations
with different signs and weights. The measured signal is a
velocity-weighted sum. One might expect this sum to wash out the Rabi
oscillations seen in cold atoms, but de Almeida and co-workers find them
preserved. The flight of atoms through the beam acts as a pulsed
excitation lasting a few microseconds, the resonant velocity groups
dominate, and the oscillations are clearer in the four-wave-mixing
signals than in transmission \cite{r28}.

Whether the switching field itself depends on velocity, and hence
whether the correlation curve in a hot vapor is in fact a sum of curves
from different velocity classes, is not settled by this work. Section 11
lists it among the open problems.

\subsection{Beyond alkali vapors}\label{sec:8.9}

Three works apply the observable outside atomic vapor.

Nadeem and co-workers extended the correlation analysis to a bulk
nitrogen--vacancy diamond crystal at 77 K, generating correlated Stokes
and anti-Stokes signals by spontaneous parametric four-wave mixing under
resonant excitation \cite{r29}. In rare-earth-doped crystals, Ahmed and
colleagues studied the switching of correlation and noise level in
Pr\textsuperscript{3+}:YSO by dressing the nonlinear phase \cite{r30}.
Wang and co-workers showed that polarization dressing controls whether a
phase-conjugate four-wave-mixing beam and a fourth-order fluorescence
beam are correlated or anticorrelated in a rare-earth-doped solid
\cite{r57}.

Solid-state emitters bring inhomogeneous broadening, dephasing mediated
by lattice vibrations (phonons), and no ground-state Zeeman structure of
the alkali type. The specific mechanism of Section 3.4 does not carry
over. What does carry over is the definition of the observable, together
with a general expectation. Two modes whose amplitudes derive from a
common, coherently driven medium tend to have correlated intensity
fluctuations, with both sign and magnitude reporting on the medium.
Three papers in three different hosts suggest a pattern, but none of
them extracts a rate from the correlation in the way Section 7 does.

\subsection{Structured light: correlated vortex generation}\label{sec:8.10}

Seth, Bhattacharyya, Mallick and De applied the observable to structured
light. They showed that the orbital angular momentum (a helical phase
structure of the beam) carried by one applied beam is transferred to a
new field generated from the prepared spin coherence in
\textsuperscript{85}Rb. One beam was locked to the D\textsubscript{1}
line to prepare the coherence and the other to the D\textsubscript{2}
line to generate the new field. They found the generated and applied
fields to be positively correlated, with extended correlation times
\cite{r31}. Combining orbital angular momentum with noise correlation adds
a spatial degree of freedom that the technique did not previously have.
The result rests on a single paper, published in 2025.

\subsection{Cavity settings and optical switching}\label{sec:8.11}

Rao, Hu, Xu and Li showed that in cavity EIT the cross-intensity
correlation changes sign as the atom--pump detunings are varied, so that
the switch is driven optically rather than magnetically \cite{r23}. Kim
and co-workers demonstrated a comparable switch from correlation to
anticorrelation for light circulating in an optical loop \cite{r58}.
Placing the medium in a cavity changes the available mode structure and
adds a second timescale, the cavity lifetime, which could eventually
make the correlation switch the basis of an all-optical switching
element. Zhou and colleagues, working with an active cavity containing a
gain medium, examined the same phase correlation at two-photon resonance
and carefully distinguished their configuration from the passive-vapor
case \cite{r47}.

\subsection{Related uses}\label{sec:8.12}

Several further works use the observable, or the underlying phenomenon,
more indirectly. Passerat de Silans, Gonçalves, Felinto and Tabosa
combined a crossover-resonance enhancement of four-wave mixing in cesium
vapor with intensity-correlation measurements \cite{r59}. Liu and
colleagues built an ultra-narrow, tunable atomic optical filter based on
EIT \cite{r60}. Xu and Hu calculated how atomic coherence, treated beyond
the adiabatic limit, shapes the spectrum of laser intensity fluctuations
under EIT \cite{r61}. Finally, the author's own later study of noise
correlations in Stokes and anti-Stokes Raman scattering from pyridine
notes that the time-domain correlation tools are not directly related to
frequency-domain two-dimensional spectroscopy. That study instead
defines the intensity--intensity noise correlation in the spectral
domain, where it equals the diagonal projection of a generalized
two-dimensional correlation map \cite{r62}.

\section{Applications and metrological consequences}\label{sec:9}

\subsection{Magnetometry}\label{sec:9.1}

The most obvious application is magnetometry, which follows directly
from the geometry: the control parameter of the switching curve
\emph{is} the magnetic field, and the curve is steepest at the zero
crossing. No such magnetometer has yet been built. If one were, it would
determine the field from the position of the correlation zero rather
than from the amplitude of a transmission or polarization-rotation
signal. It would therefore be insensitive to drifts in laser power and
to the absolute detector calibration. The same argument makes the
decoherence protocol calibration-free. Such a magnetometer would also
inherit the same weakness, since the zero lies wherever the residual
field leaves it.

Pati, Tripathi, Grewal, Yu, Krainak and Purucker demonstrated a closely
related result, the detection of magneto-optical resonances in
\textsuperscript{87}Rb at high field, proposed as a route to improved
remote magnetometry \cite{r63}. It is the closest approach to this
application in the literature, although it is a conference report. In
their account, the intensities of the two polarization channels become
correlated because the laser's own phase noise is converted into
amplitude noise on transmission, the same mechanism developed throughout
this review (Section 2.2). No full journal treatment has yet followed.
No sensitivity figure for a correlation-based magnetometer has been
published either, and it cannot be compared quantitatively with
optically pumped magnetometers \cite{r12}, which reach the femtotesla
range. Qualitatively, the trade-off is the familiar one for differential
techniques: two detection channels and a record long enough to compute a
correlation reduce the bandwidth but improve immunity to common-mode
drift.

\subsection{Clocks and the noise floor of coherent-population-trapping metrology}\label{sec:9.2}

Chip-scale CPT clocks interrogate the same ground-state coherence whose
decoherence rate the protocol of Section 7 measures, and their
short-term stability is limited in part by the same conversion of laser
frequency noise into signal noise \cite{r13}. Two consequences follow from
this overlap. Phase-to-intensity noise conversion is one of the
processes that limit the short-term stability of these devices. The
correlation properties of the Λ scheme suggest a way of identifying and
rejecting the part of that noise generated by the medium. The second
point is only a direction; no such scheme has been worked out in print.
The original stochastic treatment did name both noise reduction in
EIT-based clocks and magnetometers and decoherence metrology as
anticipated applications \cite{r16}.

Of these two, the author has found no independent application of the
decoherence protocol (Section 7.2) and no realization of the
noise-reduction idea. The clock and correlation literatures cite almost
none of the same papers. The reason is probably practical. Clock
engineering has addressed laser-noise conversion through source
stabilization rather than correlation detection, and adding a second
detection channel to a chip-scale device carries a real cost. Whether a
correlation-based rejection scheme could justify that cost remains an
open engineering question.

\subsection{Noise-limited sensitivity in Rydberg electrometry}\label{sec:9.3}

Rydberg-atom microwave electrometry reads out the microwave-induced
Autler--Townes splitting of a highly excited (Rydberg) atomic level
through electromagnetically induced transparency in a vapor cell, and
has grown into a substantial field since it was first demonstrated
\cite{r64}. Its sensitivity is limited, among other factors, by laser
frequency noise converted into intensity noise through the same
mechanism that this review describes. Yang and co-workers analyzed
exactly this problem, solving stochastic differential equations for the
atomic coherence under phase-diffusion driving to determine the
noise-limited sensitivity of a Rydberg superheterodyne receiver
\cite{r65}.

Here the physics of Section 2.2 appears as a limitation to be
quantified, although the calculation is the same as when it is used as a
resource. Any resonant sensor driven by a laser with phase noise has an
intensity-noise floor generated by the medium, and the same stochastic
methods can compute it, as Ref.~\cite{r65} does.

\subsection{Quantum memory, filters and other adjacent uses}\label{sec:9.4}

Several works cite the correlation phenomenon from within research
programs that do not use it as a measurement. Examples include Hétet's
work on continuous-variable quantum memory in atomic ensembles \cite{r48},
the EIT-based ultra-narrow atomic optical filters of Liu and colleagues
\cite{r60}. They also include the correlated photon pairs in a
rare-earth-doped solid studied by Ahmed and co-workers \cite{r30}, and the
anticorrelated switching in an optical loop reported by Kim and
colleagues \cite{r58}. What these works share is that the ground-state
coherence carrying the correlation is also the resource each application
exploits for storage, filtering or switching. Its noise properties
affect performance even where the correlation itself is not measured.

\section{How the framework has been taken up}\label{sec:10}

Not every part of the framework has spread equally. Figure 4 places the
developments of Sections 3--8 on a common timeline.

\begin{figure}[!htbp]
\centering
\includegraphics[width=0.92\linewidth]{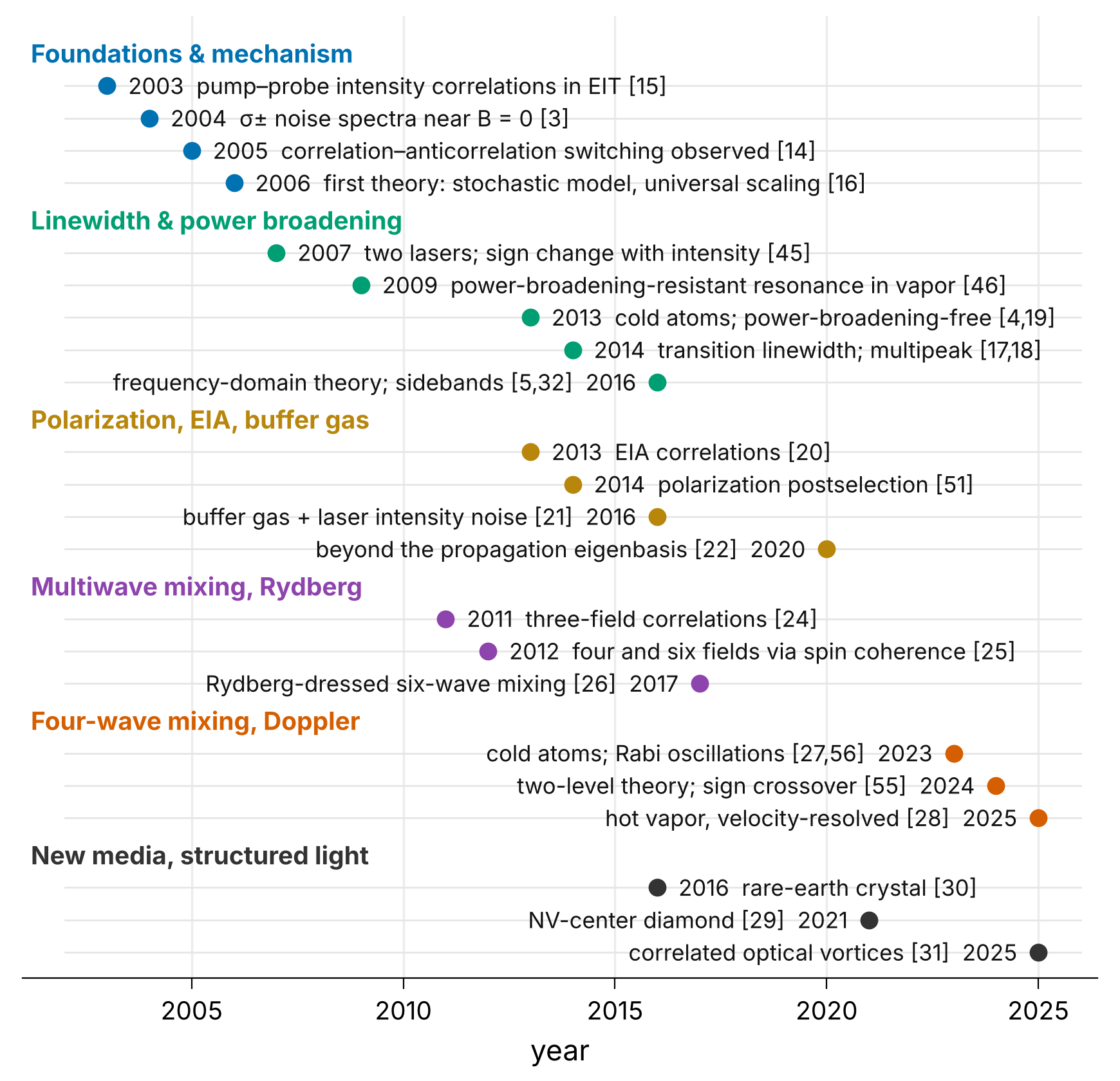}
\caption{Development of the field, 2003--2025, grouped by the physics involved. Each marker is a contribution discussed in the text, labeled with its year and reference number.}\label{fig:4}
\end{figure}

The two-mode correlation function of Section 2.3 has been adopted as a
common working definition across several lines of work. From the early
2010s onward the same normalized correlation of intensity fluctuations
appears in papers on multiwave mixing \cite{r25}, Rydberg-dressed six-wave
mixing \cite{r26}, nitrogen--vacancy diamond \cite{r29} and, most recently,
correlated vortex generation \cite{r31}. These papers usually give a
single citation, without qualification. These studies share the
observable, but not the mechanism or the theoretical model. Each
interprets the correlation within the physics of its own system. The
definition is easy to adopt because it requires nothing more than two
measurable modes.

The switching mechanism of Section 3.4 has spread differently, as
established background. It is cited as a known property of coherently
prepared Λ systems in work on cavity EIT \cite{r23}, EIA \cite{r20} and
theoretical treatments of three-level intensity correlations \cite{r33}.
It serves as the reference against which new sign structures are
compared.

The coherence readout of Section 3.2 appears wherever propagation
equations must be written down: in the Rydberg six-wave-mixing analysis
\cite{r26}, in the cold-atom EIT interpretation of Felinto and co-workers
\cite{r19}, and throughout the four-wave-mixing work \cite{r27,r28}.

The stochastic model of Section 4.1 has a narrower audience, since it is
used only where a time series must actually be computed. It acquired a
second noise channel in the buffer-gas work \cite{r21}, was complemented
by analytic perturbative treatments in the frequency-domain work
\cite{r32,r43}, and was adopted and extended as the analytical core of the
four-wave-mixing program of the 2020s \cite{r27,r28,r55}. Its most
substantial use is thus also its most recent. One interpretation,
offered here by the author and not made in the citing papers, is that a
solvable stochastic treatment becomes indispensable only once a system
is complicated enough that its correlations cannot be understood by
qualitative reasoning. Correlations between \emph{generated}
four-wave-mixing signals in a Doppler-broadened vapor are exactly such a
case.

The decoherence protocol of Section 7 and the benchmark values of
Section 5.2 have spread least of all. Figure 4 shows the pattern. The
elements that are easy to adopt have spread widely, whereas those that
require reproducing a specific experimental geometry have spread little.
Reproducing a geometry is a considerably higher bar than adopting a
definition.

\section{Conclusions}\label{sec:11}

Noise correlation spectroscopy uses the phase noise of an ordinary diode
laser to measure atomic coherence. Two beams from the same laser drive
the two arms of a Λ system and share the same frequency noise. On
two-photon resonance the transmitted beams fluctuate together. Away from
resonance, an antisymmetric fluctuation of the ground-state populations
makes them fluctuate in antiphase. The field at which the sign changes
is set by the ground-state decoherence rate, not by the optical
linewidth or the drive strength. For this reason, the correlation
resonance is several times narrower than the transparency window of the
same sample, and above a drive threshold it does not broaden with power.

The stochastic density-matrix theory introduced by the author and his
collaborators \cite{r16} describes these features with a single scale
factor and no fitting parameters. It has since gained a second noise
channel for the laser's intensity noise \cite{r21}, been complemented by
frequency-domain treatments \cite{r32}, and become the basis of
four-wave-mixing studies in cold and hot vapors \cite{r27,r28}. Experiments
have carried the technique to cold atoms, buffer-gas cells, induced
absorption, multiwave mixing, color centers, rare-earth crystals and
structured light.

The method still has clear limits. The insensitivity to the drive has
been tested over only about one decade of optical power, and its edges
have not been mapped. Residual fields, finite optical depth, velocity
averaging and laser intensity noise all reduce it. The calibration-free
measurement of the ground-state decoherence rate has not been applied
independently, or in the vapor cells used for clocks and magnetometers.
There is no closed-form theory of the switching linewidth, and the time-
and frequency-domain descriptions have not been reconciled.
Narrow-linewidth lasers, velocity-dependent switching \cite{r28} and
interacting Rydberg ensembles \cite{r26} remain unexplored. These open
problems are also the main opportunities. Once validated against
established methods, a reference-free measurement of ground-state
relaxation would be useful for compact clocks, magnetometers and quantum
memories, and could make noise correlation spectroscopy a standard
spectroscopic tool.

\section*{Funding}

This research received no external funding.

\section*{Data Availability Statement}

No new data were created in this study. All material discussed is drawn
from publicly available published literature, which is listed in the
References section.

\section*{Acknowledgments}

The author is the first author of Ref.~\cite{r16}, a co-author of Ref.~\cite{r36} and the author of Ref.~\cite{r62}. The author is not an author of
Ref.~\cite{r14}.

\section*{Conflicts of Interest}

The author declares no conflict of interest.

\end{document}